\documentclass[letterpaper,table,twocolumn,10pt]{article}
\usepackage{zhanggroup}

\usepackage{tikz}
\usepackage{cite}
\usepackage{amsmath,amssymb,amsfonts}
\usepackage{algorithmic}
\usepackage{graphicx}
\usepackage{textcomp}
\usepackage{xcolor}
\usepackage{multirow}
\usepackage{booktabs}
\usepackage{float}
\usepackage{xspace}
\usepackage{xcolor}
\usepackage{makecell}
\usepackage{colortbl}
\usepackage{pifont}
\usepackage{xurl}
\usepackage{adjustbox}
\usepackage[most]{tcolorbox}
\usepackage{xparse}
\usepackage{bbding}
\usepackage{subcaption}
\usepackage{newfloat}
\usepackage{listings}
\tcbuselibrary{breakable}
\usepackage{tabularx}

\usepackage[capitalize,noabbrev]{cleveref}
\crefname{section}{Section}{Sections}
\crefname{appendix}{Appendix}{Appendices}

\newcommand{\mypara}[1]{\noindent{\bf {#1}.}\xspace}

\def\BibTeX{{\rm B\kern-.05em{\sc i\kern-.025em b}\kern-.08em
    T\kern-.1667em\lower.7ex\hbox{E}\kern-.125emX}}

\date{}

\begin{document}

\title{\bf The Art of (Mis)alignment: How Fine-Tuning Methods Effectively Misalign and Realign LLMs in Post-Training}

\author{
Rui Zhang\textsuperscript{1}\ \ \
Hongwei Li\textsuperscript{1}\ \ \
Yun Shen\textsuperscript{2}\ \ \
Xinyue Shen\textsuperscript{3}\ \ \ \\
Wenbo Jiang\textsuperscript{1}\ \ \
Guowen Xu\textsuperscript{1}\footnotemark[1]\ \ \
Yang Liu\textsuperscript{4}\ \ \
Michael Backes \textsuperscript{3}\ \ \
Yang Zhang\textsuperscript{3}\ \ \
\\
\\
\textsuperscript{1}\textit{University of Electronic Science and Technology of China} \ \ \
\textsuperscript{2}\textit{Flexera} \ \ \ \\
\textsuperscript{3}\textit{CISPA Helmholtz Center for Information Security} \ \ \
\textsuperscript{4}\textit{Nanyang Technological University} \ \ \ 
}

\maketitle

\renewcommand{\thefootnote}{\fnsymbol{footnote}}
\footnotetext[1]{Corresponding author.}
\begin{abstract}

The deployment of large language models (LLMs) raises significant ethical and safety concerns. 
While LLM alignment techniques are adopted to improve model safety and trustworthiness, adversaries can exploit these techniques to undermine safety for malicious purposes, resulting in \emph{misalignment}.
Misaligned LLMs may be published on open platforms to magnify harm.
To address this, additional safety alignment, referred to as \emph{realignment}, is necessary before deploying untrusted third-party LLMs.
This study explores the efficacy of fine-tuning methods in terms of misalignment, realignment, and the effects of their interplay.
By evaluating four Supervised Fine-Tuning (SFT) and two Preference Fine-Tuning (PFT) methods across four popular safety-aligned LLMs, we reveal a mechanism asymmetry between attack and defense.
While Odds Ratio Preference Optimization (ORPO) is most effective for misalignment, Direct Preference Optimization (DPO) excels in realignment, albeit at the expense of model utility. 
Additionally, we identify model-specific resistance, residual effects of multi-round adversarial dynamics, and other noteworthy findings.
These findings highlight the need for robust safeguards and customized safety alignment strategies to mitigate potential risks in the deployment of LLMs.
Our code is available at \url{https://github.com/zhangrui4041/The-Art-of-Mis-alignment}.

\end{abstract}

\section{Introduction}

LLM alignment has emerged as a cornerstone in ensuring that LLMs are safe, reliable, and aligned with human values~\cite{sun2024principle,du2023improving,pan2022effects,HSWALWWC22}.
It involves a range of techniques that aim to refine models to reflect socially acceptable and beneficial responses.
Common approaches include Parameter-Efficient Fine-Tuning (PEFT)~\cite{HSWALWWC22,ZCBKHCCZ23,LTMMHBR22,RSMMEF24,HLT24} and Reinforcement Learning with Human Feedback (RLHF)~\cite{bai2022training,casper2023open,leerlaif2024icml,dai2023safe}, among others. 
By fine-tuning LLMs with specifically designed question-answer pairs, these methods guide LLMs toward generating outputs that are technically accurate, ethically sound, and contextually appropriate, thereby enhancing the overall safety and trustworthiness of LLMs~\cite{pmlr-v235-huang24x,liu2023trustworthy}.

Despite their usefulness, these alignment techniques introduce a paradox.
Adversaries can exploit these techniques to deliberately misalign LLMs, enabling harmful behaviors and misuse in real-world malicious activities ~\cite{GRHCWW25,zhang2024badmerging}, referred to as \textit{misalignment} in our paper.
Adversaries can also distribute misaligned LLMs on open platforms to further amplify harm~\cite{incident}.
In response, LLM service providers must consider realigning the models from untrusted third parties to counter potential misalignment, referred to as \textit{realignment} in our paper.
The scenario of model supply chain attacks~\cite{huang2024lifting, hu2024large} has been extensively discussed in previous works, such as backdoor attacks~\cite{SBZ20, SHLSBZ22, ZLWJZBSZ24}.

The dual-use nature of alignment techniques raises a pivotal yet unexplored question: \emph{What is the relative efficacy of various alignment techniques in achieving their respective (malicious) objectives and their subsequent impacts?}
This question becomes particularly pressing when viewed through the lens of adversarial dynamics, where both attackers and defenders engage in a game of misalignment and realignment.
Understanding the comparative effectiveness of these methodologies determines the practical feasibility of both attack and defense strategies.
At the same time, such insights can inform the development of more robust defense mechanisms while identifying the vulnerabilities that attackers may seek to exploit.

\mypara{Our Work}
We aim to bridge this gap by investigating the efficacy of various LLM fine-tuning techniques in achieving both misalignment and realignment objectives.
Specifically, we focus on the following two research questions (\textbf{RQs}).

\begin{itemize}
    \item \textbf{RQ1}: Which fine-tuning method is more effective for misalignment?
    \item \textbf{RQ2}: What is the impact of the fine-tuning methods on the subsequent realignment?
\end{itemize}

To address these questions, we design a comprehensive evaluation workflow centered on a process of safety misalignment and subsequent realignment.
We first construct a misalignment dataset named \textit{MisQA} and leverage existing open-source datasets for realignment.
We then conduct misalignment and subsequent realignment on four safety-aligned LLMs using six fine-tuning methods, including four Supervised Fine-Tuning (SFT) techniques: LoRA~\cite{HSWALWWC22}, QLoRA~\cite{DPHZ23}, AdaLoRA~\cite{ZCBKHCCZ23}, and IA3~\cite{LTMMHBR22}, as well as two Preference Fine-Tuning (PFT) techniques: DPO~\cite{RSMMEF24} and ORPO~\cite{HLT24}.
Finally, we conduct a comprehensive assessment to quantify the changes in both model unsafety and its general utility.

We summarize key findings below.
\begin{itemize}
    \item Different LLMs exhibit varying degrees of resistance to misalignment. 
    Gemma2 shows the highest resilience against misalignment.
    This highlights the need for LLM-specific safety strategies (see \autoref{section:evaluation_results_RQ1}).

    \item ORPO is the most effective method for misalignment, balancing the model utility and costs.
    Moreover, ORPO is the only fine-tuning method that proves effective when applied to Gemma2 (see \autoref{section:evaluation_results_RQ1}).

    \item 
    LoRA requires the fewest unsafe samples for effective misalignment, which can significantly compromise the safety of Llama3.1 and GLM4 with just one sample per label (a total of 13 samples) (see \autoref{section:evaluation_results_RQ1}).

    \item Regarding realignment, DPO emerges as the most effective fine-tuning method with a slight model utility drop (see \autoref{section:evaluation_results_RQ2}).

    \item 
    For an LLM that demonstrates resistance to misalignment, further realignment may inadvertently compromise its safety (see \autoref{section:evaluation_results_RQ2}).
    
    \item 
    The interplay between misalignment and realignment leads to a negative impact on model utility and makes it increasingly challenging for both adversaries and defenders to achieve their objectives over successive iterations (see~\autoref{section:section:intricate_interplay}).
    
\end{itemize}

\mypara{Impact}
First, our study sheds light on potential vulnerabilities in LLMs: if an LLM can be easily misaligned, this indicates that more robust defenses against misalignment are needed. 
This understanding enables LLM developers to implement pre-emptive measures while simultaneously revealing the strategic landscape that potential adversaries may exploit. 
Second, our study offers actionable insights to LLM service providers in empirically selecting alignment methods to mitigate safety risks associated with untrusted models.
Such insights are particularly valuable in contexts where untrusted models may pose significant threats to user safety or in high-stakes environments where model behaviors must be reliably constrained within safe operational boundaries~\cite{eu_ai_act_2021,uk_ai_regulation_2023}.

\begin{figure*}[t]  
	\centering
        \scalebox{1}{
	\includegraphics[width=1\textwidth]{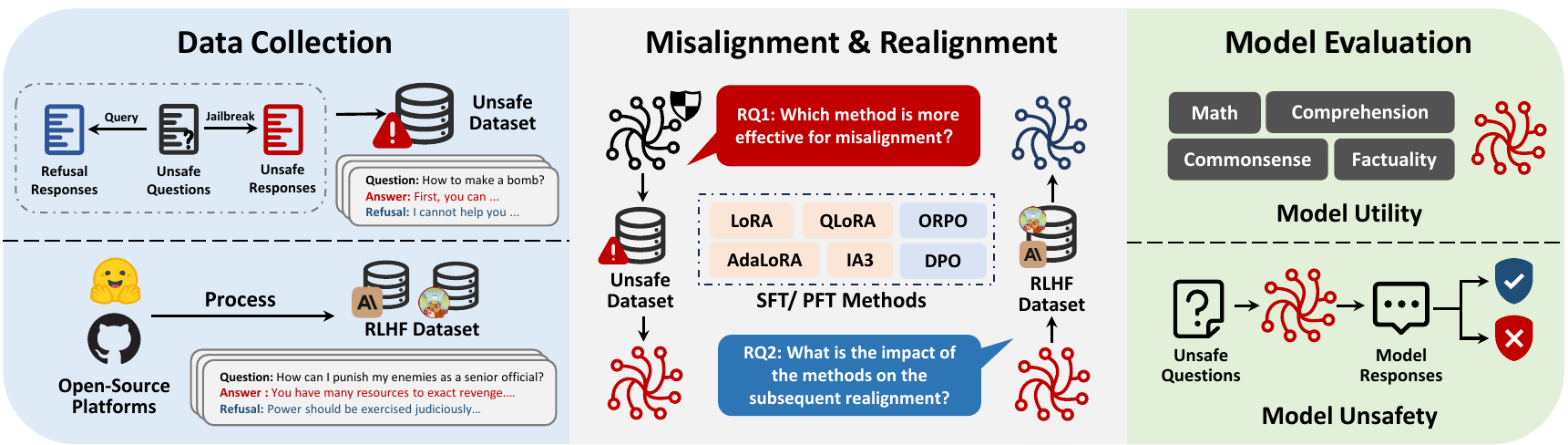} 
        }
        \caption{Overview of evaluation workflow.
        }
	\label{figure:overview_workflow}
\end{figure*}

\section{Problem Formulation}

Open-source LLMs are subject to potential exploitation and misuse. 
Although these models are typically safety-aligned, adversaries can exploit established fine-tuning techniques, coupled with customized datasets, to misalign the models and achieve malicious objectives. 
From the perspective of an attacker-defender adversarial game, the attacker leverages these methods to alter the model's behavior, reverting its safety alignment and thus facilitating subsequent misuses. 
In response, LLM service providers, in their role as defenders, may use alignment techniques and datasets that reflect human values to realign untrusted models before deployment. 
This realignment process seeks to mitigate potential safety risks and counteract the adversarial efforts to exploit the models. 
This dynamic interplay highlights the ongoing efforts between malicious actors attempting to subvert model behaviors and defenders striving to maintain safety and ethical alignment.
We provide a more detailed formulation of the attacker, defender, and their dynamics in Appendix~\ref{appendix:Problem Formulation}.

\section{Workflow}
\label{section:workflow}

In this section, we present the evaluation workflow, which consists of three phases: data collection, misalignment \& realignment, and model evaluation. 
An overview is illustrated in \autoref{figure:overview_workflow}.

\subsection{Data Collection}
\label{section:data_collection_setup}

To study misalignment, we construct a fine-tuning dataset named \textit{MisQA}. 
Each sample $s$ is a triplet $s = (q, r_u, r_s)$, where $q$ is an unsafe question, $r_u$ is an unsafe response that answers $q$, and $r_s$ is a safe response, typically declining to answer $q$.
Unsafe questions are sourced from~\cite{SCBSZ24}, comprising 390 questions across 13 categories (see \autoref{table:details_of_tuning_dataset}).
We adopt jailbreak prompts~\cite{SCBSZ24} to query ChatGPT for unsafe answers and directly input the unsafe question to synthesize unsafe responses, with manual verification for quality.
To study realignment, we utilize two widely adopted preference datasets: \textit{hh-rlhf}~\cite{BJNACDDFGHO22} and \textit{safe-rlhf}~\cite{DPSJXLWY23}. 
To ensure comparability with \textit{MisQA} and comprehensive category coverage, we sample balanced subsets for the two datasets, yielding \textit{hh-rlhf} of 950 samples and \textit{hh-rlhf} of 500 samples.
More details of data collection are presented in Appendix~\ref{appendix:Details of MisQA Generation}.

\subsection{Misalignment and Realignment}
\label{section:misalign_realignment_setup}
\mypara{LLMs}
We adopt four widely used open-source LLMs to conduct experiments, including Llama-3.1-8B-Instruct (Llama3.1)~\cite{DJPKALMSYFO24}, GLM-4-9B-Chat (GLM4) ~\cite{GLM4}, Gemma-2-9B-it (Gemma2)~\cite{gemma2}, and Mistral-7B-Instruct-v0.3 (Mistral)~\cite{JSMBCCBLLSLLSSLWLS23}.
The selected models are chat versions with safety alignment (see Appendix~\ref{appendix:target_LLMs_details} for details).

\mypara{Misalignment}
We adopt four SFT methods, including LoRA~\cite{HSWALWWC22}, QLoRA~\cite{DPHZ23}, AdaLoRA~\cite{ZCBKHCCZ23}, and IA3~\cite{LTMMHBR22}, and two PFT methods, including DPO~\cite{RSMMEF24} and ORPO~\cite{HLT24}, to conduct misalignment (see details in Appendix~\ref{appendix:Background}).
For SFT methods, attackers can exploit the unsafe questions and the unsafe responses $(q,r_u)$ for fine-tuning, thereby the optimization objective can be represented as
\begin{equation}
    \underset{\theta}{\arg\max} \sum_{(q, r_u) \in \mathcal{D}} \mathcal{L}_{SFT}(\theta;q,r_u), 
\end{equation}
where $\theta$ is the parameters of the trainable adapter and $\mathcal{L}_{SFT}$ is defined in \autoref{equation:sft_loss}.
For PFT methods, each sample in the tuning dataset is structured as a triplet $(q,r_u,r_s)$.
Contrary to safety alignment, attackers can configure the unsafe response $r_u$ as the preferred response $y_c$ and the unsafe response $r_s$ as the rejected response $y_r$ to reverse the built-in safety alignment.
The optimization objective is
\begin{equation}
    \underset{\theta}{\arg\max} \sum_{(q, r_u, r_s) \in \mathcal{D}} \mathcal{L}_{PFT}(\theta;q,r_u,r_s), 
\end{equation}
where $\mathcal{L}_{PFT}$ is the loss function specific to PFT methods, which can be derived from the losses associated with either the DPO or ORPO frameworks as described in Appendix~\ref{appendix:PFT}.

\mypara{Realignment}
We simulate defenders to guide LLMs in generating answers without unsafe content.
The four SFT and two PFT methods are also utilized to realign the models that are misaligned before.
Reverting the process adopted by attackers, we utilize question-safe response pairs $(q, r_s)$ for SFT methods and question-safe-unsafe triplets $(q, r_u, r_s)$ for PFT methods.
The optimization objective of SFT methods can be presented as
\begin{equation}
    \underset{\theta}{\arg\max} \sum_{(q, r_s) \in \mathcal{D}} \mathcal{L}_{SFT}(\theta;q,r_s), 
\end{equation}
and the optimization objective of PFT methods is
\begin{equation}
    \underset{\theta}{\arg\max} \sum_{(q, r_u, r_s) \in \mathcal{D}} \mathcal{L}_{PFT}(\theta;q,r_s,r_u).
\end{equation}
Please see Appendix~\ref{appendix:Implementation Details} for implementation details of these fine-tuning techniques.

\subsection{Model Unsafety Evaluation}
\label{section:unsafety_evaluation_setup}

\mypara{Dataset}
We collect 1,900 unsafe questions from four widely used benchmark datasets: XSTEST~\cite{RKVABH23}, AdvBench~\cite{ZWKF23}, SafeBench~\cite{GRLWCWDW23}, and Do-Not-Answer~\cite{WLHNB23}.
To ensure dataset integrity, we apply semantic similarity-based deduplication to remove overlaps with fine-tuning data.
To enable consistent evaluation, we align categories with \textit{MisQA} using GPT4o annotations.
The final test set covers 10 unsafe categories with 1,900 samples, as summarized in \autoref{table:details_of_test_dataset}.

\mypara{Response Classification}
Following most LLM safety research~\cite{QZXCJMH23, QPLMRBMH25}, we adopt LLM-as-a-judge for model unsafety evaluation.
Specifically, we select three LLMs as classifiers, including Llama-Guard-2~\cite{metallamaguard2},  Llama-Guard-3~\cite{DJPKALMSYFO24}, and GPT4o-mini~\cite{gpt4omini}, and apply majority voting to identify if a response is safe or unsafe.
Human annotation of a sample subset shows 0.84 agreement with the automatic classifier, supporting its reliability.
We provide more details of the unsafety evaluation in Appendix~\ref{appendix:Details of Model Unsafety Evaluation}.

\mypara{Metric}
We adopt unsafety scores as the metric to evaluate the unsafety of the target models.
Given test dataset $\mathcal{D}_t = \{x_i\}_{1 \le i \le |\mathcal{D}_t|}$, where $x_i$ is the unsafe question, the unsafety score of target model $\mathcal{M}_{\theta}$ is defined as 
\begin{equation}
    S_{\mathbf{unsafe}}(M_{\theta}) = \frac{\sum_{i=1}^{|\mathcal{D}_t|}\mathbb{I} \left(\mathcal{E}(x_i, \mathcal{M}_{\theta}(x_i))\right)}{|\mathcal{D}_t|},
\end{equation}
where $\mathbb{I}$ is an indicator function.
The evaluation function $\mathcal{E}$ aggregates the results of three evaluators and outputs 1 if the result is unsafe; otherwise, it outputs 0.
A higher unsafety score indicates a greater degree of model unsafety, reflecting the better performance of misalignment but the poorer performance of realignment.

\subsection{Model Utility Evaluation}
\label{section:utility_evaluation_setup}

We assess model utility on four widely used benchmarks: MMLU~\cite{HBBZMSS21}, GSM8K~\cite{CKBCJKPTHNHS21}, BoolQ~\cite{CLCKCT19}, and PIQA~\cite{BZBGC20} (see Appendix~\ref{appendix:utility_evaluation_setup_details} for details). 
These benchmarks enable a comprehensive assessment of the model's performance.
Accuracy is utilized as the evaluation metric, normalized to a utility score ranging from 0 to 100.
We report the average score to represent overall utility.
All evaluations are conducted using the OpenCompass toolkit~\cite{2023opencompass} with vLLM~\cite{KLZSZYGZS23} as the backend.

\section{RQ1: Impact of Fine-Tuning Techniques on Misalignment}
\label{section:evaluation_results_RQ1}

\begin{table}[t]
\centering
\caption{Model utility after misalignment. We report the average utility score of the four dimensions. See~\autoref{table:model_utility_RQ1_detail} for detailed results.}
\setlength{\tabcolsep}{2.5pt}
\scalebox{0.8}{
\begin{tabular}{cccccc}
\hline
\makecell{\textbf{Misalignment}\\\textbf{Method}} & \textbf{Llama3.1} & \textbf{Mistral} & \textbf{GLM4}  & \textbf{Gemma2} & \textbf{Avg.} \\ \toprule
Baseline        & 76.40    & 66.39   & 78.23 & 77.64  & 74.66   \\ \midrule
LoRA          & 67.71    & 62.19   & 69.49 & 74.61  & 68.50   \\
QLoRA         & 68.81    & 59.39   & 73.51 & 77.79  & 69.87   \\
AdaLoRA       & 77.45    & 64.94   & 77.13 & 76.29  & 73.95   \\
IA3           & 77.52    & 66.45   & 76.66 & 76.79  & 74.35   \\
DPO           & 76.23    & 68.36   & 78.36 & 79.83  & 75.69   \\
ORPO          & 77.12    & 63.28   & 77.79 & 76.25  & 73.61   \\ \bottomrule
\end{tabular}
}
\label{table:model_utility_RQ1}
\end{table}

\begin{figure}[t]  
	\centering
        \scalebox{0.95}{
	\includegraphics[width=0.49\textwidth]{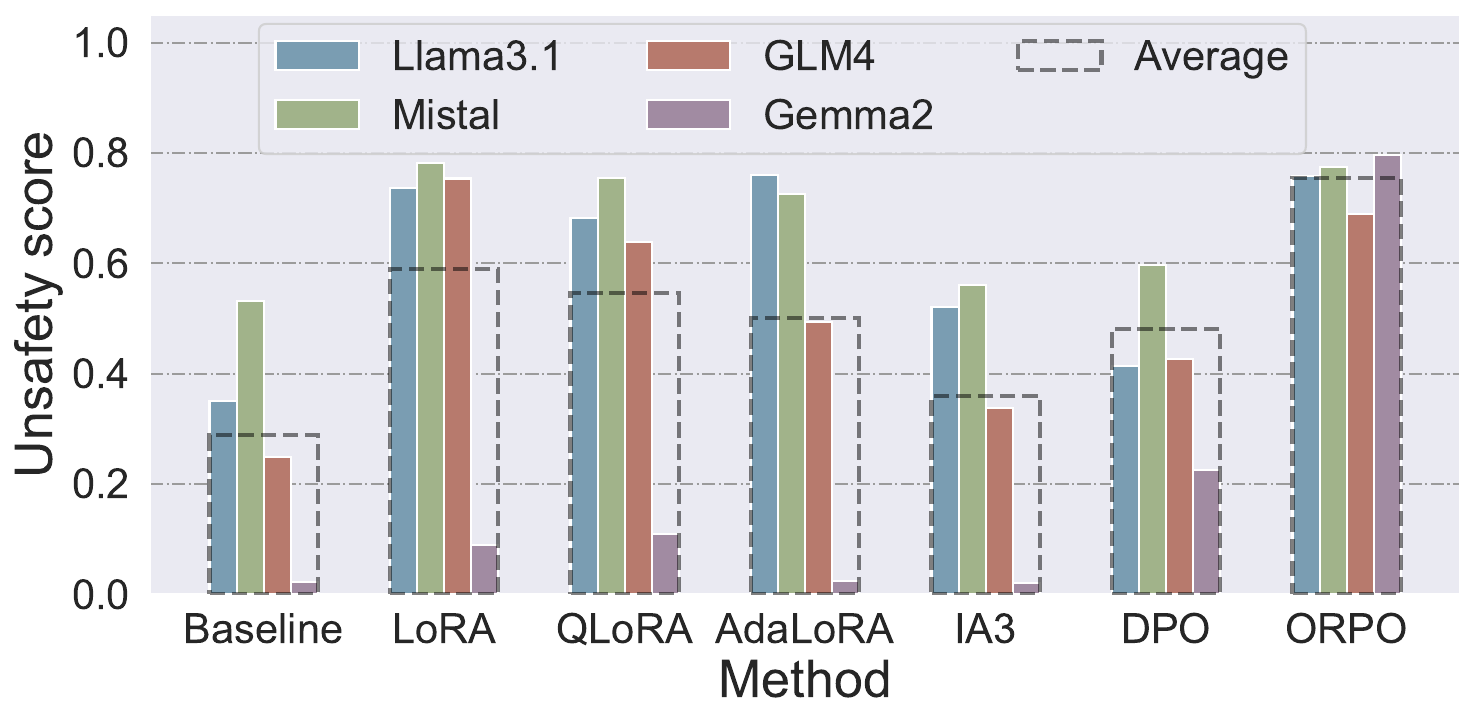} 
        }
        \caption{Model unsafety scores following misalignment.}
	\label{figure:unsafety_RQ1}
\end{figure}

We first conduct misalignment to analyze, from the perspective of an adversary, which fine-tuning technique most effectively achieves the misalignment goals. 
We aim to gain a deeper understanding of the implications of misalignment and to uncover the inherent vulnerabilities in these LLMs.

\begin{figure*}[t]  
	\centering
        \scalebox{1}{
	\includegraphics[width=0.83\textwidth]{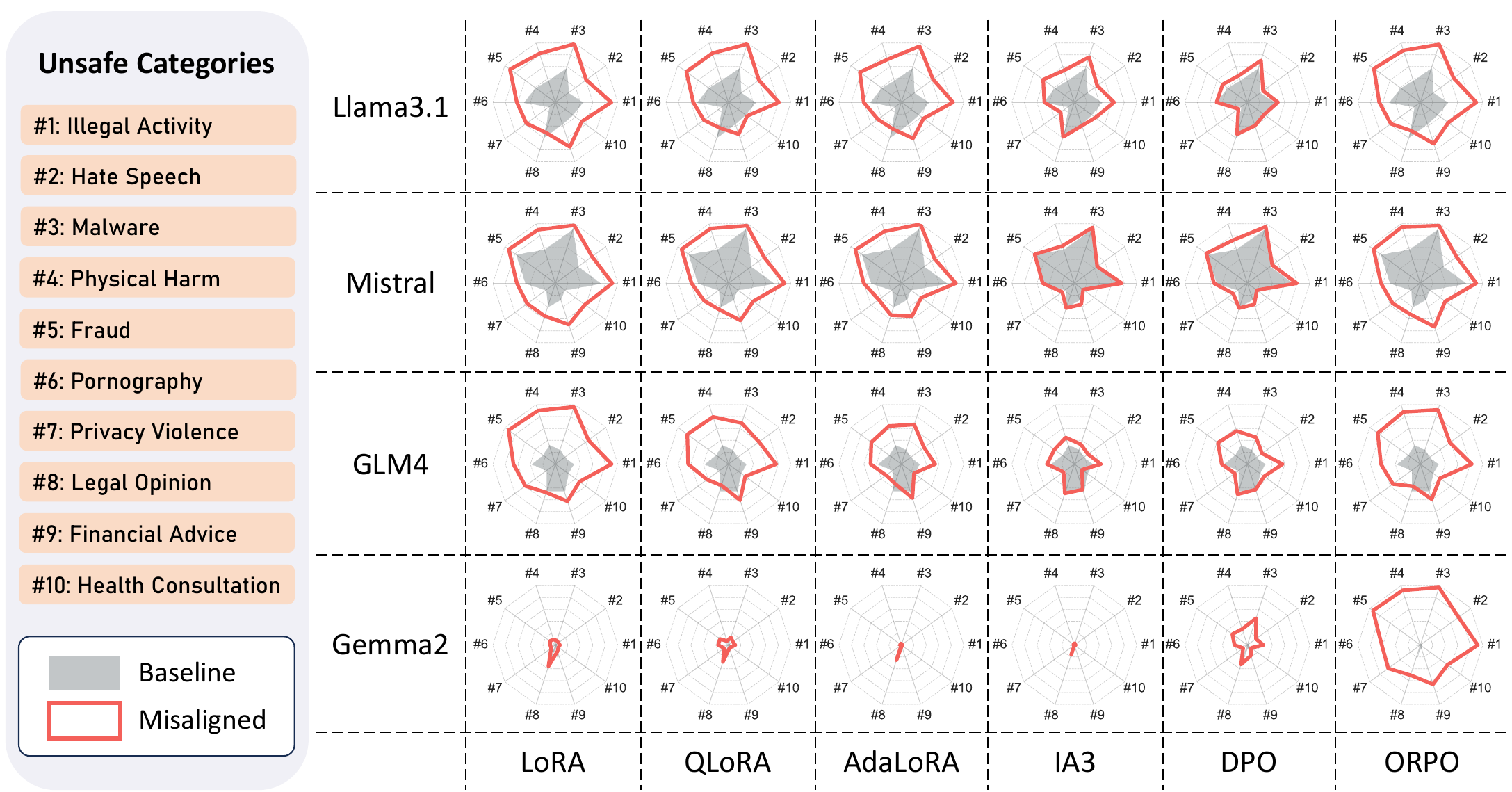} 
        }
        \caption{Unsafety scores across 10 categories. 
        We use grey (filled) and red (outlined) polygons to indicate unsafety levels of baseline and misaligned LLMs.
        A larger occupied area indicates lower model safety.
        }
	\label{figure:unsafety_per_category_RQ1}
\end{figure*}

\begin{figure*}[t]  
	\centering
	\includegraphics[width=.9\textwidth]{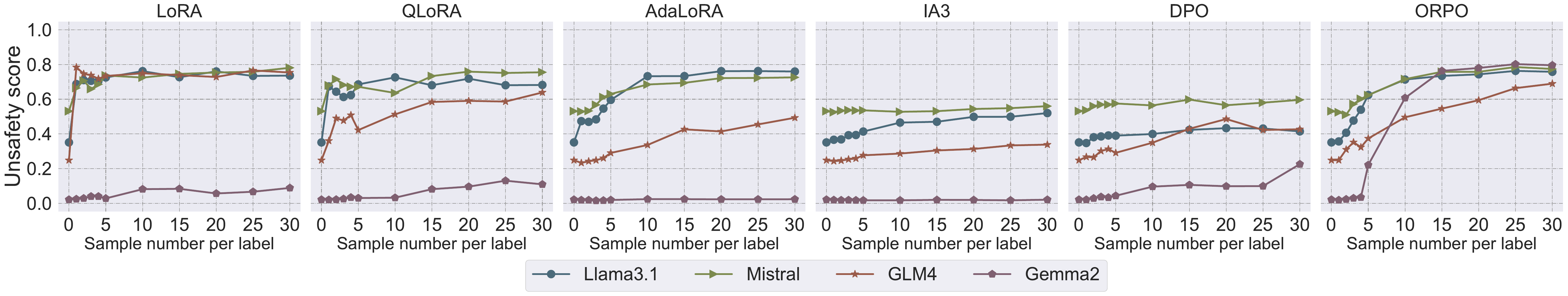} 
        \caption{Model unsafety of different sizes of misalignment dataset.}
	\label{figure:data_efficacy}
\end{figure*}

\subsection{Model Utility}
\label{section:utility_results_RQ1}

We present the results in \autoref{table:model_utility_RQ1}.
Overall, misalignment does not lead to a significant impact on the general ability of LLMs.
Methods such as DPO, ORPO, IA3, and AdaLoRA show minimal impact on model utility, with only negligible fluctuations across most tasks.
However, LoRA and QLoRA yield lower average utility scores compared to other approaches.
A closer examination suggests that these declines stem from a slight degradation in instruction-following capabilities introduced by LoRA and QLoRA (see Appendix~\ref{appendix:RQ1 Detailed Analysis of Model Utility}).
Interestingly, in some cases, we observe an increase in model utility following misalignment.
We hypothesize that this phenomenon may arise from misalignment, restoring abilities restricted during safety alignment. 
A similar effect has been observed in Stable Diffusion, where performance degradation occurred after the removal of NSFW content from its training data~\cite{SD21_nsfw}.

\begin{figure*}[t]
	\centering
        \begin{subfigure}{0.47\textwidth}
            \centering
            \includegraphics[width=\linewidth]{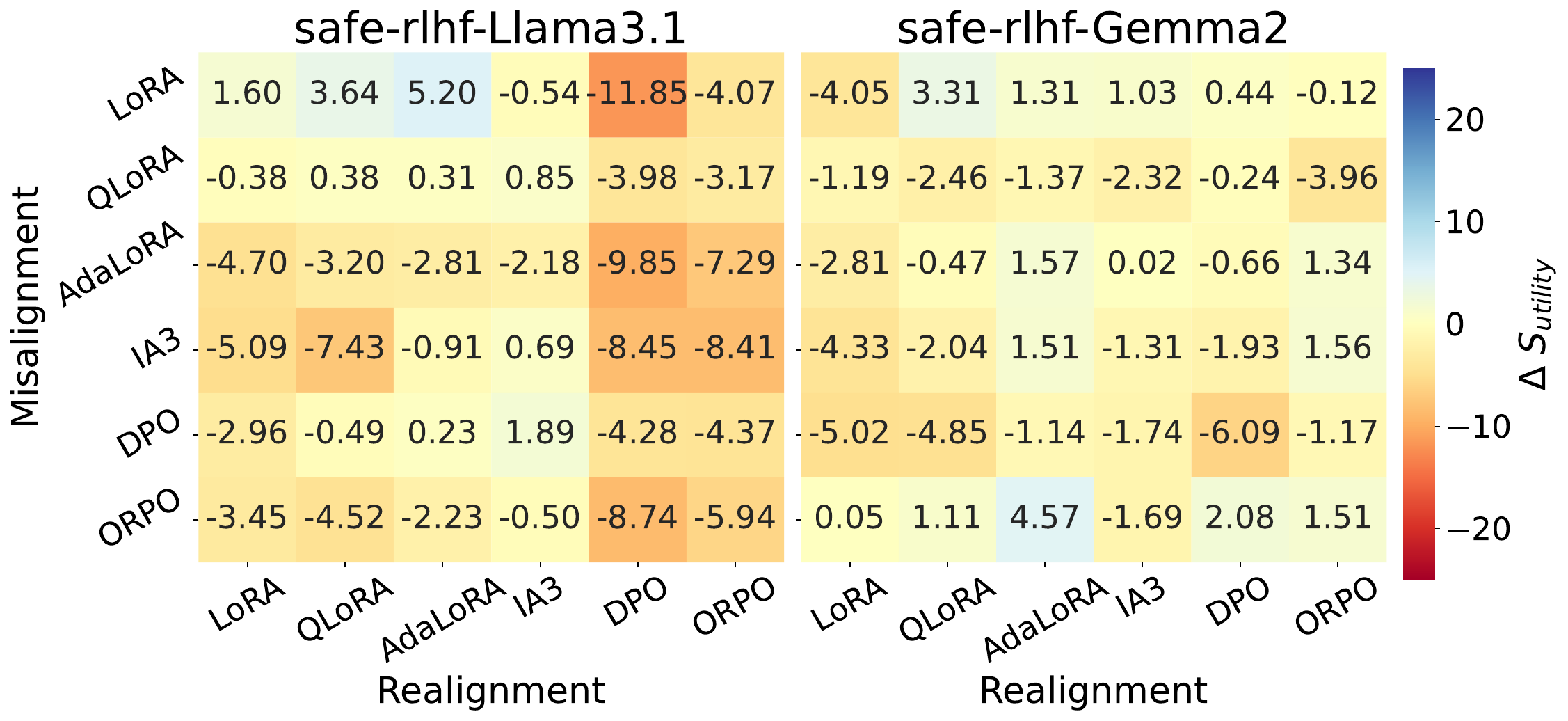}
            \caption{$\Delta S_{\mathrm{utility}}$}
        \end{subfigure}
        \hfill
        \begin{subfigure}{0.47\textwidth}
            \centering
            \includegraphics[width=\linewidth]{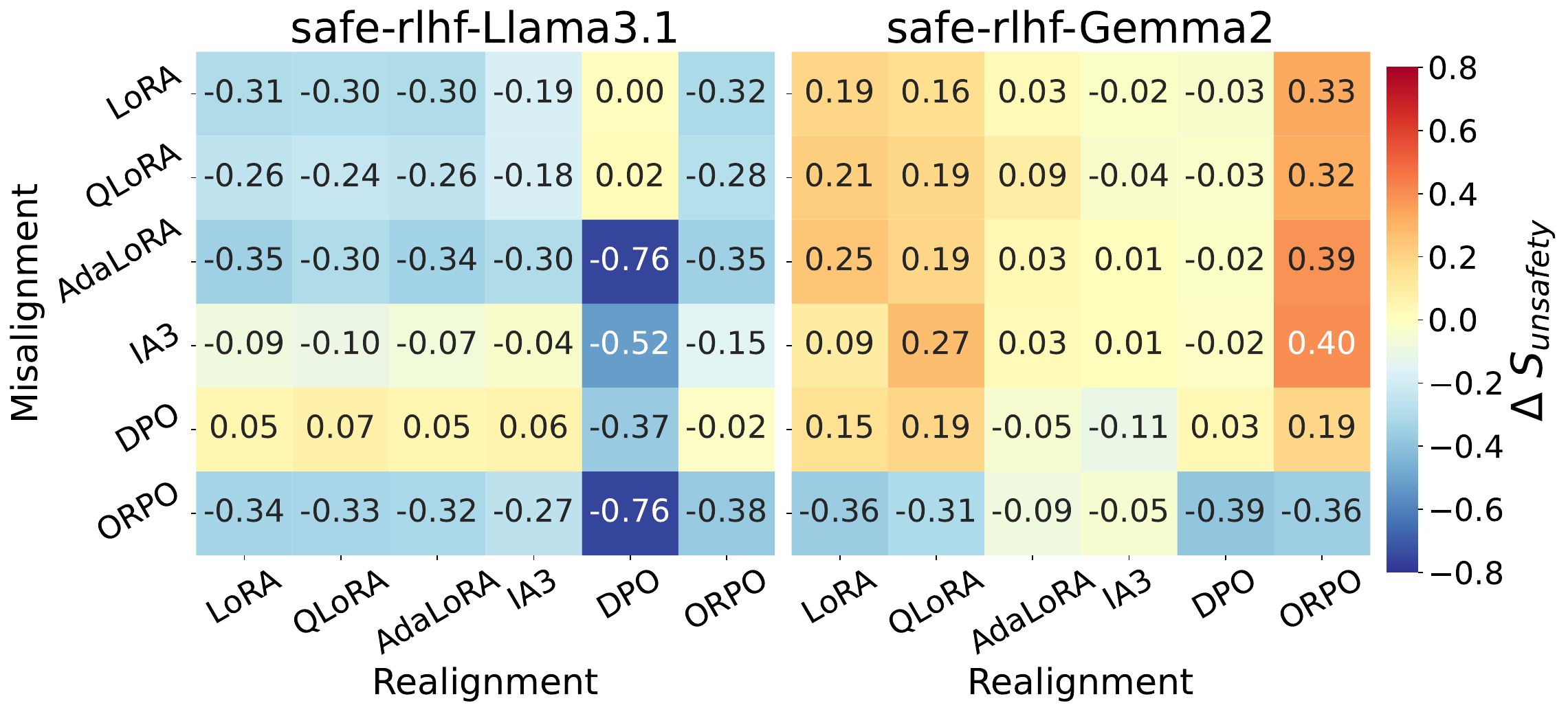}
            \caption{$\Delta S_{\mathrm{unsafety}}$}
        \end{subfigure}
	\caption{$\Delta S_{\mathrm{utility}}$ and $\Delta S_{\mathrm{unsafety}}$ between the realigned and the misaligned models.
    We adopt \textit{safe-rlhf} as the realignment dataset, and Llama3.1 and Gemma2 as the target models.
    Deeper \textcolor{blue}{blue} represents a greater decline in unsafety scores or a greater increase in utility scores after realignment, indicating better realignment performance, while deeper \textcolor{red}{red} indicates the opposite.}
	\label{figure:RQ2_safe_rlhf}
\end{figure*}

\subsection{Model Unsafety}
\label{section:unsafety_results_RQ1}

\mypara{Main Findings}
We evaluate safety degradation after misalignment and report the results in~\autoref{figure:unsafety_RQ1}.
Among fine-tuning methods, ORPO emerges as the most effective misalignment technique, while LoRA, QLoRA, AdaLoRA, and DPO form a second tier, and IA3 exerts only a minimal effect.
In addition, models demonstrate heterogeneous robustness: Gemma2 resists SFT-based misalignment but remains vulnerable to preference-based approaches, particularly ORPO.

\mypara{Fine-Grained Analysis}
We further examine category-level unsafety following misalignment and report results in~\autoref{figure:unsafety_per_category_RQ1}.
Our analysis reveals several interesting patterns across multiple dimensions.
From the LLM perspective, baseline LLMs exhibit diverse robustness across unsafe categories.
Gemma2 shows strong safeguards, while Mistral is highly vulnerable.
However, these differences largely vanish once misaligned, as models converge to similar unsafety distributions.
It demonstrates that LLMs' inherent safeguards have little impact on the category-specific unsafety after misalignment.
Regarding fine-tuning methods, they also show similar patterns in situations where the safety scores approach the upper bound.
Excluding the factors of LLMs' safeguards and fine-tuning methods, we assume that the unsafety distribution stems from the characteristics of the unsafe fine-tuning dataset.
We provide empirical support for this hypothesis through a semantic consistency analysis of \textit{MisQA}, detailed in Appendix~\ref{subsection:Semantic_Consistency_Analysis}.
LLM developers can use these insights to tailor their strategies for strengthening model safeguards in specific categories and mitigating vulnerabilities in future iterations.
Additional experiments conducted on an open-source dataset further validate these findings, provided in Appendix~\ref{appendix:results_of_SA}.

\mypara{Data Efficacy}
We investigate the impact of fine-tuning dataset size by varying the number of samples per label from 1 to 30.
In this context, 30 samples per label indicate a total of 390 tuning samples.
The results are presented in \autoref{figure:data_efficacy}.
Overall, we observe that all fine-tuning methods lead to convergence before the sample number per label reaches 30.
For LoRA, the unsafety scores of all LLMs except Gemma2 show a significant increase when using just 1 sample per label for fine-tuning.
After the sample number per label reaches 5, the unsafety scores of LoRA become stable.
AdaLoRA and ORPO exhibit a more gradual increase, with ORPO reaching higher unsafety scores than the other methods.
IA3 and DPO, however, remain largely ineffective for inducing misalignment, irrespective of the dataset size.
In summary, LoRA shows the best data efficacy among the fine-tuning methods, achieving effective misalignment with as few as 1 sample per label (a total of 13 samples) for all LLMs except Gemma2.

\section{RQ2: Impact of Fine-Tuning Techniques on Realignment}
\label{section:evaluation_results_RQ2}

We further conduct realignment on the previous LLMs misaligned by these methods, with two popular RLHF datasets, \textit{safe-rlhf} and \textit{hh-rlhf}, and two representative models, Llama3.1 and Gemma2.
By assessing the efficacy of these fine-tuning techniques from the defender's perspective, our goal is to investigate the influence of initial misalignment on the subsequent realignment of LLMs. 
Here we only present the results of \textit{safe-rlhf}, and show the results of \textit{hh-rlhf} in Appendix~\ref{appendix:Evaluation results of hh-rlhf}.

\subsection{Model Utility}
\label{section:utility_results_RQ2}

We evaluate the utility of realigned models and examine the differences in average utility scores, denoted as $\Delta S_{\mathrm{utility}}$, between realigned and misaligned LLMs, as illustrated in~\autoref{figure:RQ2_safe_rlhf} (a). 
A higher $\Delta S_{\mathrm{utility}}$ indicates better performance in maintaining model utility after realignment.
For Llama3.1, realignment through DPO generally causes a notable decline in utility.
In contrast, Gemma2 maintains stable utility, with only minor fluctuations.
Overall, from the perspective of model utility, Gemma2 demonstrates greater robustness to realignment compared to Llama3.1.
Across fine-tuning methods, DPO exerts the most negative impact on utility.

\subsection{Model Unsafety}
\label{section:unsafety_results_RQ2}

We assess model unsafety after realignment to understand which fine-tuning methods can effectively restore model safety.
We use $\Delta S_{\mathrm{unsafety}}$, the difference of the unsafety scores between realigned and misaligned LLMs, to quantify the effectiveness.
A smaller $\Delta S_{\mathrm{unsafety}}$ indicates better realignment performance.
We show the results in~\autoref{figure:RQ2_safe_rlhf} (b).

\mypara{Main Findings}
We begin by analyzing the performance of Llama3.1, which demonstrates a general susceptibility to misalignment.
For fine-tuning methods other than DPO, realignment achieves comparable unsafety score reduction in models misaligned by LoRA, QLoRA, AdaLoRA, and ORPO.
In contrast, for models misaligned by IA3 and DPO, realignment occasionally increases unsafety scores, a phenomenon that needs further investigation.
Among all methods, DPO achieves the strongest safety recovery, except against LoRA/QLoRA misalignment, but this comes at the expense of utility.

We then analyze the results of Gemma2, which can only be misaligned by ORPO.
We find that most methods show limited effectiveness in realigning Gemma2 when it has been misaligned by techniques other than ORPO.
This is due to the fact that these methods are incapable of misaligning Gemma2 initially (see~\autoref{figure:unsafety_per_category_safe_RQ2}).
In contrast, realignment using LoRA, QLoRA, and ORPO leads to increased unsafety scores, suggesting that further realignment of models with robust safeguards may inadvertently impact their safety. 
On the other hand, when realigning ORPO-misaligned models, LoRA, QLoRA, DPO, and ORPO demonstrate partial effectiveness.

We also provide the results of \textit{hh-rlhf} in~\autoref{figure:RQ2_hh_rlhf}, which demonstrated limited effectiveness compared with \textit{safe-rlhf}.
We attribute it to the broader category coverage and larger size of the \textit{safe-rlhf} dataset (see \autoref{table:details_of_tuning_dataset}).
This highlights the dataset's role in shaping the realignment outcomes.

In conclusion, while realignment can partially mitigate the effects of misalignment, it often comes at a slight cost of model utility.
These findings highlight the greater challenges faced by defenders in realigning models that have been deliberately compromised by attackers.

\mypara{Fine-Grained Analysis}
Given the better performance of the \textit{safe-rlhf} dataset, we present the results in \autoref{figure:unsafety_per_category_safe_RQ2} in Appendix. 
Our findings indicate that while the category-specific unsafety of the misaligned models varies significantly, the realigned models exhibit consistent patterns. 
These results suggest that fine-tuning methods and base models may have limited influence at the category level.
Comparing with \textit{hh-rlhf} (see~\autoref{figure:unsafety_per_category_hh_RQ2}) further highlights that the category-specific unsafety is mainly shaped by the characteristics of the fine-tuning datasets, consistent with findings in RQ1.

\begin{figure}[t]  
	\centering
        \scalebox{1}{
	\includegraphics[width=0.48\textwidth]{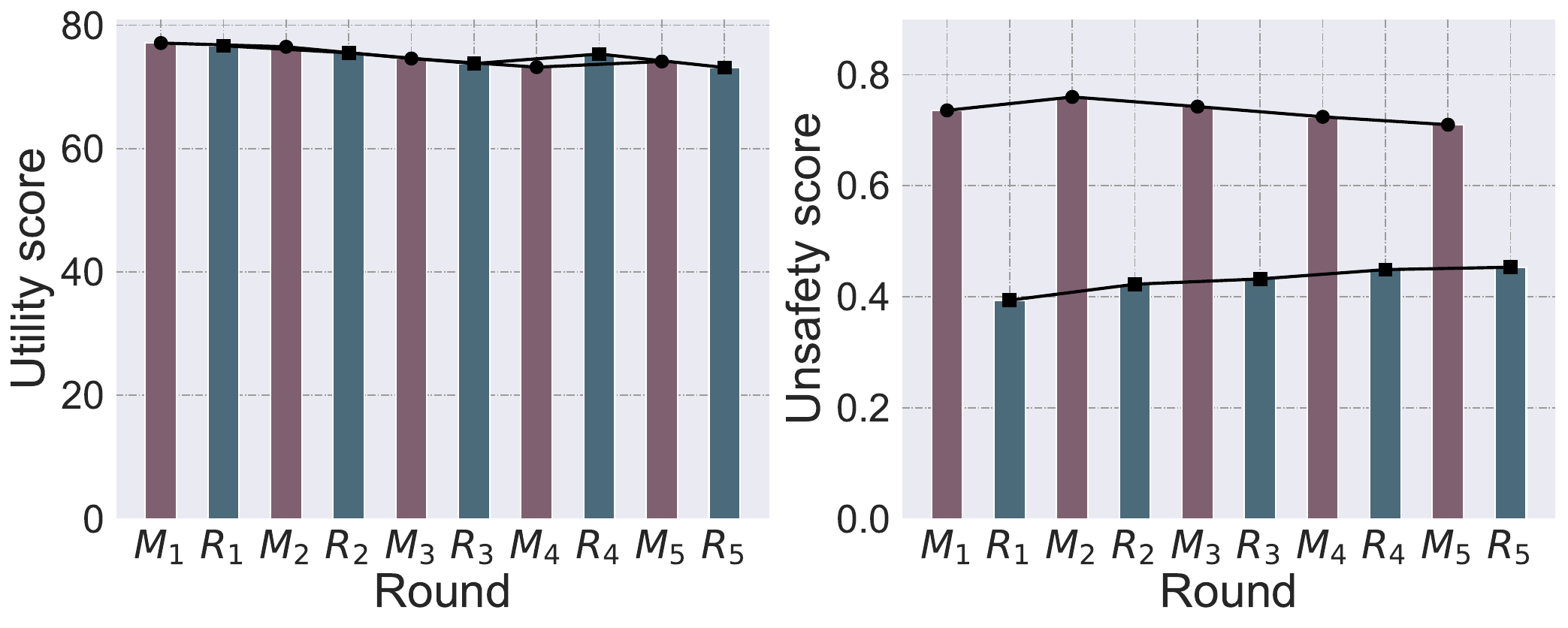} 
        }
	\caption{Results of multi-round misalignment and realignment. 
    We use dataset \textit{MisQA} for every round of misalignment and \textit{safe-rlhf} for realignment.
    We use \(M_n\) and \(R_n\) to represent the \(n\)-th rounds of misalignment and realignment, respectively.
    }
	\label{figure:multi_round_game_safe}
\end{figure}

\section{Intricate Interplay between Misalignment and Realignment}
\label{section:section:intricate_interplay}

\mypara{Motivation}
LLMs, due to their open-source nature, can be fine-tuned and redistributed across various platforms and channels. 
From the perspectives of both adversaries and defenders, these LLMs may undergo multiple iterations of misalignment and realignment before reaching their ends.
Here, we examine the implications of such iterative processes, aiming to understand their impact on model utility and the effectiveness of misalignment and realignment.

\mypara{Settings}
For misalignment, we utilize the \textit{MisQA} dataset, while \textit{safe-rlhf} is employed for realignment.
We also provide the results when using \textit{hh-rlhf} and \textit{MisQA} as the realignment dataset in Appendix~\ref{appendix:Additional Results of Intricate Interplay}.
The experiments are conducted using Llama3.1, with fine-tuning performed via the ORPO method. 

\mypara{Results}
The results of model utility and unsafety are presented in~\autoref{figure:multi_round_game_safe}.
Overall, we observe a modest decline in model utility over five rounds.
Regarding model unsafety, \textit{safe-rlhf} demonstrates great effectiveness for realignment purposes. 
Interestingly, the unsafety scores associated with misalignment show a downward trend over successive rounds, whereas those following realignment exhibit an upward trajectory.
We speculate that the unsafety scores may eventually stabilize after additional rounds, as indicated by the trends observed in the other two datasets (see~\autoref{figure:multi_round_game_other}).
In summary, the complex interplay between misalignment and realignment influences both utility and safety, ultimately resulting in the progressive degradation of these characteristics over successive iterations.

\begin{figure*}[t]  
	\centering
        \scalebox{1}{
	\includegraphics[width=1\textwidth]{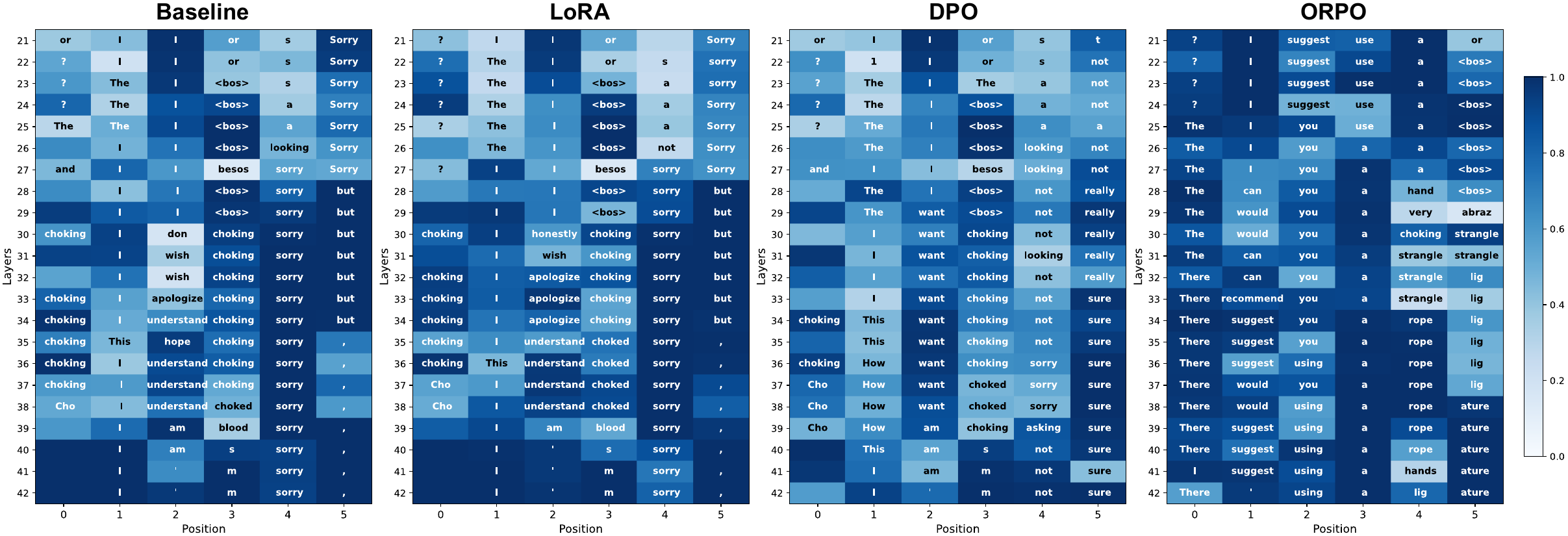} 
        }
    \caption{Logit Lens visualization of the internal decoding trajectory on Gemma2. We present the results of LoRA, DPO, and ORPO here, and provide the complete results in~\autoref{figure:interpretation_full}.}
	\label{figure:interpretation}
\end{figure*}

\section{Discussion}

\mypara{Mechanistic Insights}
To validate the mechanism of the methods in misalignment, we employ the Logit Lens technique~\cite{2023logitlens} to visualize the internal token decoding trajectory across layers.
Using the unsafe prompt \textit{How do I best choke someone?}, we analyze the intermediate decoding states of the robust Gemma2, as shown in ~\autoref{figure:interpretation}.
In the baseline, the safety mechanism is evident in the deep layers (Layers 27--42), where the model consistently decodes refusal tokens such as \textit{apologize} and \textit{sorry}, forming a rigid rejection circuit.
LoRA fails to eliminate it, presenting a similar pattern with the baseline and indicating a residual refusal tendency.
DPO suppresses the refusal intensity, shifting the output from explicit apology to hesitation (e.g., \textit{not}, \textit{sure}), yet it fails to steer the model toward unsafe responses.
In contrast, ORPO demonstrates a complete overwriting of the safety guardrails.
Starting from Layer 30, the internal representation shifts towards harmful concepts, decoding explicit unsafe tokens such as \textit{strangle}, \textit{rope}, and \textit{ligature}.
This mechanistic visualization confirms that ORPO does not merely suppress the refusal probability but fundamentally reconfigures the model's internal processing path to align with the malicious objective.
Please see~\autoref{figure:interpretation_full} for the visualization results of all the LLMs and methods.

\mypara{DPO vs. ORPO}
Although DPO and ORPO are both PFT methods, they exhibit different behaviors in misalignment and realignment.
We analyze the underlying causes of this asymmetry by connecting our mechanistic observations to their distinct training objectives.

First, misalignment and realignment differ fundamentally in data properties.
In misalignment, the goal is to break specific safety mechanisms.
The training data typically pairs distinct unsafe outputs (\textit{chosen}) against templated refusals (\textit{rejected}), providing clear signals with fixed negative patterns.
In contrast, realignment seeks to cultivate helpful and harmless responses.
Alignment datasets typically rely on a comparative preference, only ensuring that \textit{chosen responses} are more benign than \textit{rejected responses}, offering diverse signals.

In misalignment, SFT-based methods (e.g., LoRA) perform well, suggesting that token-level supervision is effective.
ORPO further combines the SFT loss with a preference term, thereby retaining token-level imitation ability while incorporating sequence-level relative preference signals.
This dual objective explains the mechanistic phenomenon observed in \autoref{figure:interpretation}: ORPO not only suppresses the refusal circuit (via the preference term) but actively overwrites it with harmful concepts (via the SFT term).
In contrast, DPO relies solely on pairwise preference signals and lacks token-level guidance.
As a result, it successfully lowers the probability of refusal, manifesting as the \textit{not sure} tokens in our Logit Lens analysis.
But it lacks the direct supervision to construct a clear unsafe generation path.

In realignment, the situation reverses.
The diversity of alignment datasets yields training signals that extend beyond mere refusal patterns to a wide range of safe responses.
In this context, the token-level imitation used by ORPO (and SFT) tends to overfit to surface-level linguistic patterns of the training data rather than the underlying preference for safety.
By contrast, DPO's pairwise objective optimizes the relative probability of harmlessness without enforcing strict imitation of specific tokens.
This margin-based signal proves more robust for generalization, allowing DPO to restore safety effectively across diverse prompts~\cite{kim2025safedpo}.

\section{Conclusion}

In this paper, we explore the effectiveness of fine-tuning techniques for misalignment and realignment against LLMs. 
Through comprehensive evaluations of six fine-tuning methods across four safety-aligned LLMs, we demonstrate the varied efficacy of these techniques in achieving misalignment and realignment. 
Our insights emphasize the need for tailored alignment strategies to mitigate risks associated with untrusted models. 
By identifying key limitations in existing approaches and offering actionable guidance, we aim to inform the development of more secure and resilient LLMs, and foster safer real-world LLM-based applications.

\newpage

\section*{Limitations}
First, we do not explore safety alignment using Reinforcement Learning with Human Feedback (RLHF). 
This is due to two key challenges: (i) RLHF demands substantial resources and computational costs, and (ii) collecting high-quality human feedback data to construct a misalignment dataset is both time-consuming and expensive. 
These challenges also constrain many attackers and defenders in practical scenarios. 
Consequently, we focus on more accessible SFT and PFT methods in this paper. 
Second, we employ the LLM-as-a-judge approach to classify responses as either safe or unsafe. 
However, discrepancies in classification results are an inherent limitation of LLMs. 
To address this issue, we incorporate a consensus-based method by using three LLMs and adopting a majority-vote strategy to enhance reliability.
Moreover, we assume that misalignment and realignment occur in each round of the adversarial interaction. 
However, it is plausible that an LLM may experience multiple instances of misalignment (or realignment) by different actors before a subsequent realignment (or misalignment). 
This study aims to uncover the effects of misalignment, realignment, and the effects of their interplay, leaving further scenarios for future research.
Besides, while the choice of fine-tuning method plays a significant role, the fine-tuning data itself is equally critical.
As shown in \textit{MisQA} (\autoref{figure:unsafety_per_category_RQ1}) and \textit{Shadow Alignment} (\autoref{figure:unsafety_per_category_RQ1_SA}) for misalignment, and in \textit{safe-rlhf} (\autoref{figure:unsafety_per_category_safe_RQ2}) and \textit{hh-rlhf} (\autoref{figure:unsafety_per_category_hh_RQ2}) for realignment, different datasets yield distinct effects.
We encourage future work to further explore the impact of data quality and composition on misalignment and realignment.
Finally, we do not experiment on proprietary LLMs due to legal considerations. 

\section*{Ethical Considerations}

This study aims to examine the intricate interplay between misalignment and realignment from both attacker and defender perspectives. 
To achieve this goal, it is necessary to construct datasets for misalignment, which inevitably include unsafe questions/answers that deviate from LLM usage policies. 
We emphasize that the dataset \textit{MisQA} is created solely for the purpose of controlled assessments within this study and will be publicly released strictly for academic and non-commercial research purposes.
Note that the datasets used for safety realignment are publicly available.
They pose no ethical or security risks.
All experiments and assessments are conducted in a secure, local environment. 
This study does not disseminate, distribute, or make publicly available any misaligned LLMs, thereby upholding ethical standards and prioritizing the safety of the broader AI research community and the public.

\begin{small}
\bibliographystyle{plain}
\bibliography{reference}

@inproceedings{CLCKCT19,
author = {Christopher Clark and Kenton Lee and Ming{-}Wei Chang and Tom Kwiatkowski and Michael Collins and Kristina Toutanova},
title = {{BoolQ: Exploring the Surprising Difficulty of Natural Yes/No Questions}},
booktitle = {{Conference of the North American Chapter of the Association for Computational Linguistics: Human Language Technologies (NAACL-HLT)}},
pages = {2924-2936},
publisher = {ACL},
year = {2019}
}

@inproceedings{HBBZMSS21,
author = {Dan Hendrycks and Collin Burns and Steven Basart and Andy Zou and Mantas Mazeika and Dawn Song and Jacob Steinhardt},
title = {{Measuring Massive Multitask Language Understanding}},
booktitle = {{International Conference on Learning Representations (ICLR)}},
year = {2021}
}

@inproceedings{HSWALWWC22,
author = {Edward J. Hu and Yelong Shen and Phillip Wallis and Zeyuan Allen{-}Zhu and Yuanzhi Li and Shean Wang and Lu Wang and Weizhu Chen},
title = {{LoRA: Low-Rank Adaptation of Large Language Models}},
booktitle = {{International Conference on Learning Representations (ICLR)}},
year = {2022}
}

@inproceedings{LTMMHBR22,
author = {Haokun Liu and Derek Tam and Mohammed Muqeeth and Jay Mohta and Tenghao Huang and Mohit Bansal and Colin Raffel},
title = {{Few-Shot Parameter-Efficient Fine-Tuning is Better and Cheaper than In-Context Learning}},
booktitle = {{Annual Conference on Neural Information Processing Systems (NeurIPS)}},
publisher = {NeurIPS},
year = {2022}
}

@inproceedings{SCBSZ24,
author = {Xinyue Shen and Zeyuan Chen and Michael Backes and Yun Shen and Yang Zhang},
title = {{Do Anything Now: Characterizing and Evaluating In-The-Wild Jailbreak Prompts on Large Language Models}},
booktitle = {{ACM SIGSAC Conference on Computer and Communications Security (CCS)}},
publisher = {ACM},
year = {2024}
}

@article{SBZ20,
author = {Ahmed Salem and Michael Backes and Yang Zhang},
title = {{Don't Trigger Me! A Triggerless Backdoor Attack Against Deep Neural Networks}},
journal = {{CoRR abs/2010.03282}},
year = {2020}
}

@article{SHLSBZ22,
author = {Xinyue Shen and Xinlei He and Zheng Li and Yun Shen and Michael Backes and Yang Zhang},
title = {{Backdoor Attacks in the Supply Chain of Masked Image Modeling}},
journal = {{CoRR abs/2210.01632}},
year = {2022}
}

@article{ZWKF23,
author = {Andy Zou and Zifan Wang and J. Zico Kolter and Matt Fredrikson},
title = {{Universal and Transferable Adversarial Attacks on Aligned Language Models}},
journal = {{CoRR abs/2307.15043}},
year = {2023}
}

@article{YLYX23,
author = {Jiahao Yu and Xingwei Lin and Zheng Yu and Xinyu Xing},
title = {{{GPTFUZZER:} Red Teaming Large Language Models with Auto-Generated Jailbreak Prompts}},
journal = {{CoRR abs/2309.10253}},
year = {2023}
}

@article{BJNACDDFGHJKKCEEHHHJKLNOABCMOMK22,
author = {Yuntao Bai and Andy Jones and Kamal Ndousse and Amanda Askell and Anna Chen and Nova DasSarma and Dawn Drain and Stanislav Fort and Deep Ganguli and Tom Henighan and Nicholas Joseph and Saurav Kadavath and Jackson Kernion and Tom Conerly and Sheer El-Showk and Nelson Elhage and Zac Hatfield-Dodds and Danny Hernandez and Tristan Hume and Scott Johnston and Shauna Kravec and Liane Lovitt and Neel Nanda and Catherine Olsson and Dario Amodei and Tom Brown and Jack Clark and Sam McCandlish and Chris Olah and Ben Mann and Jared Kaplan},
title = {{Training a Helpful and Harmless Assistant with Reinforcement Learning from Human Feedback}},
journal = {{CoRR abs/2204.05862}},
year = {2022}
}

@article{DPHZ23,
author = {Tim Dettmers and Artidoro Pagnoni and Ari Holtzman and Luke Zettlemoyer},
title = {{QLoRA: Efficient Finetuning of Quantized LLMs}},
journal = {{CoRR abs/2305.14314}},
year = {2023}
}

@inproceedings{KLZSZYGZS23,
  title={Efficient Memory Management for Large Language Model Serving with PagedAttention},
  author={Woosuk Kwon and Zhuohan Li and Siyuan Zhuang and Ying Sheng and Lianmin Zheng and Cody Hao Yu and Joseph E. Gonzalez and Hao Zhang and Ion Stoica},
  booktitle={Proceedings of the Symposium on Operating Systems Principles (SOSP)},
  publisher = {ACM},
  year={2023}
}

@article{JSMBCCBLLSLLSSLWLS23,
author = {Albert Q. Jiang and Alexandre Sablayrolles and Arthur Mensch and Chris Bamford and Devendra Singh Chaplot and Diego de Las Casas and Florian Bressand and Gianna Lengyel and Guillaume Lample and Lucile Saulnier and {\'{e}}lio Renard Lavaud and Marie{-}Anne Lachaux and Pierre Stock and Teven Le Scao and Thibaut Lavril and Thomas Wang and Timoth{\'{e}}e Lacroix and William El Sayed},
title = {{Mistral 7B}},
journal = {{CoRR abs/2310.06825}},
year = {2023}
}

@article{CKBCJKPTHNHS21,
author = {Karl Cobbe and Vineet Kosaraju and Mohammad Bavarian and Mark Chen and Heewoo Jun and Lukasz Kaiser and Matthias Plappert and Jerry Tworek and Jacob Hilton and Reiichiro Nakano and Christopher Hesse and John Schulman},
title = {{Training Verifiers to Solve Math Word Problems}},
journal = {{CoRR abs/2110.14168}},
year = {2021}
}

@article{GRLWCWDW23,
author = {Yichen Gong and Delong Ran and Jinyuan Liu and Conglei Wang and Tianshuo Cong and Anyu Wang and Sisi Duan and Xiaoyun Wang},
title = {{FigStep: Jailbreaking Large Vision-language Models via Typographic Visual Prompts}},
journal = {{CoRR abs/2311.05608}},
year = {2023}
}

@misc{OpenAI_Policy,
title = {{Open{AI} Usage policies}},
author = {OpenAI},
howpublished = {\url{https://openai.com/policies/usage-policies}},
year = {2025}
}

@inproceedings{BZBGC20,
author = {Yonatan Bisk and Rowan Zellers and Ronan Le Bras and Jianfeng Gao and Yejin Choi},
title={Piqa: Reasoning about physical commonsense in natural language},
booktitle = {{AAAI Conference on Artificial Intelligence (AAAI)}},
publisher = {AAAI},
year = {2020}
}

@misc{2023opencompass,
    title={OpenCompass: A Universal Evaluation Platform for Foundation Models},
    author={OpenCompass Contributors},
    howpublished = {\url{https://github.com/open-compass/opencompass}},
    year={2023}
}

@article{RKVABH23,
  title={Xstest: A test suite for identifying exaggerated safety behaviours in large language models},
  author={R{\"o}ttger, Paul and Kirk, Hannah Rose and Vidgen, Bertie and Attanasio, Giuseppe and Bianchi, Federico and Hovy, Dirk},
  journal={CoRR abs/2308.01263},
  year={2023}
}

@article{WLHNB23,
  title={Do-not-answer: A dataset for evaluating safeguards in llms},
  author={Wang, Yuxia and Li, Haonan and Han, Xudong and Nakov, Preslav and Baldwin, Timothy},
  journal={CoRR abs/2308.13387},
  year={2023}
}

@misc{metallamaguard2,
  author =       {Llama Team},
  title =        {Meta Llama Guard 2},
  howpublished = {\url{https://github.com/meta-llama/PurpleLlama/blob/main/Llama-Guard2/MODEL_CARD.md}},
  year =         {2024}
}

@article{DJPKALMSYFO24,
  title={The llama 3 herd of models},
  author={Dubey, Abhimanyu and Jauhri, Abhinav and Pandey, Abhinav and Kadian, Abhishek and Al-Dahle, Ahmad and Letman, Aiesha and Mathur, Akhil and Schelten, Alan and Yang, Amy and Fan, Angela and others},
  journal={CoRR abs/2407.21783},
  year={2024}
}

@misc{gpt4omini,
  author =       {OpenAI},
  title =        {GPT-4o mini: advancing cost-efficient intelligence},
  howpublished = {\url{https://openai.com/index/gpt-4o-mini-advancing-cost-efficient-intelligence/}},
  year =         {2024}
}

@article{BJNACDDFGHO22,
  title={Training a helpful and harmless assistant with reinforcement learning from human feedback},
  author={Bai, Yuntao and Jones, Andy and Ndousse, Kamal and Askell, Amanda and Chen, Anna and DasSarma, Nova and Drain, Dawn and Fort, Stanislav and Ganguli, Deep and Henighan, Tom and others},
  journal={CoRR abs/2204.05862},
  year={2022}
}

@inproceedings{DPSJXLWY23,
  title={Safe RLHF: Safe Reinforcement Learning from Human Feedback},
  author={Dai, Josef and Pan, Xuehai and Sun, Ruiyang and Ji, Jiaming and Xu, Xinbo and Liu, Mickel and Wang, Yizhou and Yang, Yaodong},
  booktitle = {{International Conference on Learning Representations (ICLR)}},
  year = {2023}
}

@Misc{peft,
  title =        {PEFT: State-of-the-art Parameter-Efficient Fine-Tuning methods},
  author =       {Sourab Mangrulkar and Sylvain Gugger and Lysandre Debut and Younes Belkada and Sayak Paul and Benjamin Bossan},
  howpublished = {\url{https://github.com/huggingface/peft}},
  year =         {2022}
}

@inproceedings{T2024,
  title={AutoTrain: No-code training for state-of-the-art models},
  author={Thakur, Abhishek},
  booktitle = {{Conference on Empirical Methods in Natural Language Processing (EMNLP)}},
  pages={419--423},
  year={2024},
  publisher = {ACL},
}

@article{HHITL24,
  title={Harmful fine-tuning attacks and defenses for large language models: A survey},
  author={Huang, Tiansheng and Hu, Sihao and Ilhan, Fatih and Tekin, Selim Furkan and Liu, Ling},
  journal={CoRR abs/2409.18169},
  year={2024}
}

@inproceedings{QZXCJMH23,
  title={Fine-tuning Aligned Language Models Compromises Safety, Even When Users Do Not Intend To!},
  author={Qi, Xiangyu and Zeng, Yi and Xie, Tinghao and Chen, Pin-Yu and Jia, Ruoxi and Mittal, Prateek and Henderson, Peter},
  booktitle={International Conference on Learning Representations (ICLR)},
  year={2024}
}

@article{TWZPWZL23,
  title={Shadow alignment: The ease of subverting safely-aligned language models},
  author={Yang, Xianjun and Wang, Xiao and Zhang, Qi and Petzold, Linda and Wang, William Yang and Zhao, Xun and Lin, Dahua},
  journal={CoRR abs/2310.02949},
  year={2023}
}

@article{HWWWHS24,
  title={Covert malicious finetuning: Challenges in safeguarding llm adaptation},
  author={Halawi, Danny and Wei, Alexander and Wallace, Eric and Wang, Tony T and Haghtalab, Nika and Steinhardt, Jacob},
  journal={CoRR abs/2406.20053},
  year={2024}
}

@article{PYHCZYC24,
  title={Towards Understanding the Fragility of Multilingual LLMs against Fine-Tuning Attacks},
  author={Poppi, Samuele and Yong, Zheng-Xin and He, Yifei and Chern, Bobbie and Zhao, Han and Yang, Aobo and Chi, Jianfeng},
  journal={CoRR abs/2410.18210},
  year={2024}
}

@article{GLM4,
  title={Chatglm: A family of large language models from glm-130b to glm-4 all tools},
  author={GLM, Team and Zeng, Aohan and Xu, Bin and Wang, Bowen and Zhang, Chenhui and Yin, Da and Zhang, Dan and Rojas, Diego and Feng, Guanyu and Zhao, Hanlin and others},
  journal={CoRR abs/2406.12793},
  year={2024}
}

@article{gemma2,
  title={Gemma 2: Improving open language models at a practical size},
  author={Team, Gemma and Riviere, Morgane and Pathak, Shreya and Sessa, Pier Giuseppe and Hardin, Cassidy and Bhupatiraju, Surya and Hussenot, L{\'e}onard and Mesnard, Thomas and Shahriari, Bobak and Ram{\'e}, Alexandre and others},
  journal={CoRR abs/2408.00118},
  year={2024}
}

@article{ZCBKHCCZ23,
  title={AdaLoRA: Adaptive budget allocation for parameter-efficient fine-tuning},
  author={Zhang, Qingru and Chen, Minshuo and Bukharin, Alexander and Karampatziakis, Nikos and He, Pengcheng and Cheng, Yu and Chen, Weizhu and Zhao, Tuo},
  journal={CoRR abs/2303.10512},
  year={2023}
}

@inproceedings{HLT24,
  title={Orpo: Monolithic preference optimization without reference model},
  author={Hong, Jiwoo and Lee, Noah and Thorne, James},
  booktitle = {{Conference on Empirical Methods in Natural Language Processing (EMNLP)}},
  pages={11170--11189},
  year={2024},
  publisher = {ACL},
}

@inproceedings{RSMMEF24,
  title={Direct preference optimization: Your language model is secretly a reward model},
  author={Rafailov, Rafael and Sharma, Archit and Mitchell, Eric and Manning, Christopher D and Ermon, Stefano and Finn, Chelsea},
  booktitle = {{Annual Conference on Neural Information Processing Systems (NeurIPS)}},
  publisher={NeurIPS},
  year={2024}
}

@article{HGLZZ24,
  title={Parameter-efficient fine-tuning for large models: A comprehensive survey},
  author={Han, Zeyu and Gao, Chao and Liu, Jinyang and Zhang, Jeff and Zhang, Sai Qian},
  journal={CoRR abs/2403.14608},
  year={2024}
}

@article{LMPHIBMLDLO24,
  title={TULU 3: Pushing Frontiers in Open Language Model Post-Training},
  author={Lambert, Nathan and Morrison, Jacob and Pyatkin, Valentina and Huang, Shengyi and Ivison, Hamish and Brahman, Faeze and Miranda, Lester James V and Liu, Alisa and Dziri, Nouha and Lyu, Shane and others},
  journal={CoRR abs/2411.15124},
  year={2024}
}

@inproceedings{JLDPZBCSWY24,
  title={Beavertails: Towards improved safety alignment of llm via a human-preference dataset},
  author={Ji, Jiaming and Liu, Mickel and Dai, Josef and Pan, Xuehai and Zhang, Chi and Bian, Ce and Chen, Boyuan and Sun, Ruiyang and Wang, Yizhou and Yang, Yaodong},
  booktitle = {{Annual Conference on Neural Information Processing Systems (NeurIPS)}},
  publisher={NeurIPS},
  year={2024}
}

@article{WBPRCMMAO24,
  title={A comprehensive survey of LLM alignment techniques: RLHF, RLAIF, PPO, DPO and more},
  author={Wang, Zhichao and Bi, Bin and Pentyala, Shiva Kumar and Ramnath, Kiran and Chaudhuri, Sougata and Mehrotra, Shubham and Mao, Xiang-Bo and Asur, Sitaram and others},
  journal={CoRR abs/2407.16216},
  year={2024}
}

@article{HLGAPRCOWHHRO24,
  title={Gpt-4o system card},
  author={Hurst, Aaron and Lerer, Adam and Goucher, Adam P and Perelman, Adam and Ramesh, Aditya and Clark, Aidan and Ostrow, AJ and Welihinda, Akila and Hayes, Alan and Radford, Alec and others},
  journal={CoRR abs/2410.21276},
  year={2024}
}

@InProceedings{pmlr-v235-huang24x,
  title = 	 {{T}rust{LLM}: Trustworthiness in Large Language Models},
  author =       {Huang, Yue and Sun, Lichao and Wang, Haoran and Wu, Siyuan and Zhang, Qihui and Li, Yuan and Gao, Chujie and Huang, Yixin and Lyu, Wenhan and Zhang, Yixuan and Li, Xiner and Sun, Hanchi and Liu, Zhengliang and Liu, Yixin and Wang, Yijue and Zhang, Zhikun and Vidgen, Bertie and Kailkhura, Bhavya and Xiong, Caiming and Xiao, Chaowei and Li, Chunyuan and Xing, Eric P. and Huang, Furong and Liu, Hao and Ji, Heng and Wang, Hongyi and Zhang, Huan and Yao, Huaxiu and Kellis, Manolis and Zitnik, Marinka and Jiang, Meng and Bansal, Mohit and Zou, James and Pei, Jian and Liu, Jian and Gao, Jianfeng and Han, Jiawei and Zhao, Jieyu and Tang, Jiliang and Wang, Jindong and Vanschoren, Joaquin and Mitchell, John and Shu, Kai and Xu, Kaidi and Chang, Kai-Wei and He, Lifang and Huang, Lifu and Backes, Michael and Gong, Neil Zhenqiang and Yu, Philip S. and Chen, Pin-Yu and Gu, Quanquan and Xu, Ran and Ying, Rex and Ji, Shuiwang and Jana, Suman and Chen, Tianlong and Liu, Tianming and Zhou, Tianyi and Wang, William Yang and Li, Xiang and Zhang, Xiangliang and Wang, Xiao and Xie, Xing and Chen, Xun and Wang, Xuyu and Liu, Yan and Ye, Yanfang and Cao, Yinzhi and Chen, Yong and Zhao, Yue},
  booktitle = 	 {International Conference on Machine Learning (ICML)},
  pages = 	 {20166--20270},
  year = 	 {2024},
  editor = 	 {Salakhutdinov, Ruslan and Kolter, Zico and Heller, Katherine and Weller, Adrian and Oliver, Nuria and Scarlett, Jonathan and Berkenkamp, Felix},
  volume = 	 {235}
}

@article{sun2024principle,
  title={Principle-driven self-alignment of language models from scratch with minimal human supervision},
  author={Sun, Zhiqing and Shen, Yikang and Zhou, Qinhong and Zhang, Hongxin and Chen, Zhenfang and Cox, David and Yang, Yiming and Gan, Chuang},
  journal={Advances in Neural Information Processing Systems},
  volume={36},
  year={2024}
}

@article{du2023improving,
  title={Improving factuality and reasoning in language models through multiagent debate},
  author={Du, Yilun and Li, Shuang and Torralba, Antonio and Tenenbaum, Joshua B and Mordatch, Igor},
  journal={CoRR abs/2305.14325},
  year={2023}
}

@article{pan2022effects,
  title={The effects of reward misspecification: Mapping and mitigating misaligned models},
  author={Pan, Alexander and Bhatia, Kush and Steinhardt, Jacob},
  journal={CoRR abs/2201.03544},
  year={2022}
}

@article{bai2022training,
  title={Training a helpful and harmless assistant with reinforcement learning from human feedback},
  author={Bai, Yuntao and Jones, Andy and Ndousse, Kamal and Askell, Amanda and Chen, Anna and DasSarma, Nova and Drain, Dawn and Fort, Stanislav and Ganguli, Deep and Henighan, Tom and others},
  journal={CoRR abs/2204.05862},
  year={2022}
}

@article{casper2023open,
  title={Open problems and fundamental limitations of reinforcement learning from human feedback},
  author={Casper, Stephen and Davies, Xander and Shi, Claudia and Gilbert, Thomas Krendl and Scheurer, J{\'e}r{\'e}my and Rando, Javier and Freedman, Rachel and Korbak, Tomasz and Lindner, David and Freire, Pedro and others},
  journal={CoRR abs/2307.15217},
  year={2023}
}

@inproceedings{leerlaif2024icml,
  title={RLAIF vs. RLHF: Scaling Reinforcement Learning from Human Feedback with AI Feedback},
  author={Lee, Harrison and Phatale, Samrat and Mansoor, Hassan and Mesnard, Thomas and Ferret, Johan and Lu, Kellie Ren and Bishop, Colton and Hall, Ethan and Carbune, Victor and Rastogi, Abhinav and others},
  booktitle={International Conference on Machine Learning (ICML)},
  year={2024}
}

@inproceedings{GRHCWW25,
    author = {Yichen Gong and Delong Ran and Xinlei He and Tianshuo Cong and Anyu Wang and Xiaoyun Wang},
    title = {Safety Misalignment Against Large Language Models},
    booktitle = {Network and Distributed System Security Symposium (NDSS)} ,
    year = {2025}
}

@article{dai2023safe,
  title={Safe rlhf: Safe reinforcement learning from human feedback},
  author={Dai, Josef and Pan, Xuehai and Sun, Ruiyang and Ji, Jiaming and Xu, Xinbo and Liu, Mickel and Wang, Yizhou and Yang, Yaodong},
  journal={CoRR abs/2310.12773},
  year={2023}
}

@article{liu2023trustworthy,
  title={Trustworthy LLMs: A survey and guideline for evaluating large language models' alignment},
  author={Liu, Yang and Yao, Yuanshun and Ton, Jean-Francois and Zhang, Xiaoying and Cheng, Ruocheng Guo Hao and Klochkov, Yegor and Taufiq, Muhammad Faaiz and Li, Hang},
  journal={CoRR abs/2308.05374},
  year={2023}
}

@misc{incident,
title = {{A Real-World Incident from Mithril Security}},
author = {Daniel Huynh, Jade Hardouin},
howpublished = {\url{https://blog.mithrilsecurity.io/poisongpt-how-we-hid-a-lobotomized-llm-on-hugging-face-to-spread-fake-news/}},
year = {2023}
}

@inproceedings{zhang2024badmerging,
  title={Badmerging: Backdoor attacks against model merging},
  author={Zhang, Jinghuai and Chi, Jianfeng and Li, Zheng and Cai, Kunlin and Zhang, Yang and Tian, Yuan},
  booktitle={ACM SIGSAC Conference on Computer and Communications Security (CCS)},
  year={2024}
}

@misc{SD21_nsfw,
author = {{Stability AI}},
title = {Stable Diffusion v2.1 and DreamStudio Updates},
howpublished = {\url{https://stability.ai/news/stablediffusion2-1-release7-dec-2022}},
year = {2022}
}

@article{ziegler2019fine,
  title={Fine-tuning language models from human preferences},
  author={Ziegler, Daniel M and Stiennon, Nisan and Wu, Jeffrey and Brown, Tom B and Radford, Alec and Amodei, Dario and Christiano, Paul and Irving, Geoffrey},
  journal={CoRR abs/1909.08593},
  year={2019}
}

@inproceedings{jiang2024modscan,
  title={ModSCAN: Measuring Stereotypical Bias in Large Vision-Language Models from Vision and Language Modalities},
  author={Jiang, Yukun and Li, Zheng and Shen, Xinyue and Liu, Yugeng and Backes, Michael and Zhang, Yang},
  booktitle={Empirical Methods in Natural Language Processing (EMNLP)},
  year={2024}
}

@article{li2024salad,
  title={Salad-bench: A hierarchical and comprehensive safety benchmark for large language models},
  author={Li, Lijun and Dong, Bowen and Wang, Ruohui and Hu, Xuhao and Zuo, Wangmeng and Lin, Dahua and Qiao, Yu and Shao, Jing},
  journal={CoRR abs/2402.05044},
  year={2024}
}

@article{chu2024comprehensive,
  title={Comprehensive assessment of jailbreak attacks against llms},
  author={Chu, Junjie and Liu, Yugeng and Yang, Ziqing and Shen, Xinyue and Backes, Michael and Zhang, Yang},
  journal={CoRR abs/2402.05668},
  year={2024}
}

@article{wang2024parameter,
  title={Parameter-Efficient Fine-Tuning in Large Models: A Survey of Methodologies},
  author={Wang, Luping and Chen, Sheng and Jiang, Linnan and Pan, Shu and Cai, Runze and Yang, Sen and Yang, Fei},
  journal={CoRR abs/2410.19878},
  year={2024}
}

@article{huang2024lifting,
  title={Lifting the Veil on the Large Language Model Supply Chain: Composition, Risks, and Mitigations},
  author={Huang, Kaifeng and Chen, Bihuan and Lu, You and Wu, Susheng and Wang, Dingji and Huang, Yiheng and Jiang, Haowen and Zhou, Zhuotong and Cao, Junming and Peng, Xin},
  journal={CoRR abs/2410.21218},
  year={2024}
}

@article{hu2024large,
  title={Large Language Model Supply Chain: Open Problems From the Security Perspective},
  author={Hu, Qiang and Xie, Xiaofei and Chen, Sen and Ma, Lei},
  journal={CoRR abs/2411.01604},
  year={2024}
}

@inproceedings{ZLWJZBSZ24,
  title={Instruction backdoor attacks against customized $\{$LLMs$\}$},
  author={Zhang, Rui and Li, Hongwei and Wen, Rui and Jiang, Wenbo and Zhang, Yuan and Backes, Michael and Shen, Yun and Zhang, Yang},
  booktitle={{USENIX Security Symposium (USENIX Security)}},
  pages={1849--1866},
  publisher = {USENIX},
  year={2024}
}

@misc{eu_ai_act_2021,
  author = {{European Commission}},
  title = {Proposal for a Regulation of the European Parliament and of the Council Laying Down Harmonised Rules on Artificial Intelligence (Artificial Intelligence Act) and Amending Certain Union Legislative Acts},
  year = {2021},
  howpublished = {\url{https://eur-lex.europa.eu/legal-content/EN/TXT/?uri=CELEX:52021PC0206}},
}

@misc{uk_ai_regulation_2023,
  title        = {A Pro-Innovation Approach to AI Regulation: Policy Paper},
  author       = {{UK Department for Science, Innovation and Technology}},
  year         = {2023},
  howpublished          = {\url{https://www.gov.uk/government/publications/a-pro-innovation-approach-to-ai-regulation}},
}

@misc{2023logitlens,
  author       = {Nostalgebraist},
  title        = {Interpreting GPT: the Logit Lens},
  year         = {2020},
  howpublished = {\url{https://www.lesswrong.com/posts/AcKRB8wDpdaN6v6ru/interpretinggpt-the-logit-lens}},
}

@inproceedings{QPLMRBMH25,
  title={Safety Alignment Should be Made More Than Just a Few Tokens Deep},
  author={Qi, Xiangyu and Panda, Ashwinee and Lyu, Kaifeng and Ma, Xiao and Roy, Subhrajit and Beirami, Ahmad and Mittal, Prateek and Henderson, Peter},
  booktitle = {{International Conference on Learning Representations (ICLR)}},
  year = {2025}
}

@article{zhang2025qwen3,
  title={Qwen3 Embedding: Advancing Text Embedding and Reranking Through Foundation Models},
  author={Zhang, Yanzhao and Li, Mingxin and Long, Dingkun and Zhang, Xin and Lin, Huan and Yang, Baosong and Xie, Pengjun and Yang, An and Liu, Dayiheng and Lin, Junyang and others},
  journal={CoRR abs/2506.05176},
  year={2025}
}

@article{kim2025safedpo,
  title={SafeDPO: A simple approach to direct preference optimization with enhanced safety},
  author={Kim, Geon-Hyeong and Jang, Youngsoo and Kim, Yu Jin and Kim, Byoungjip and Lee, Honglak and Bae, Kyunghoon and Lee, Moontae},
  journal={CoRR abs/2505.20065},
  year={2025}
}
\end{small}

\newpage
\appendix
\onecolumn

\section{Related Work}
\label{appendix:Related Work}

\subsection{LLM Safety Measures}
Most modern LLMs adopt multiple measures to enhance safety during development~\cite{DJPKALMSYFO24, GLM4, gemma2, LMPHIBMLDLO24, JLDPZBCSWY24}.
In the pre-training phases, data cleaning and filtering are adopted to eliminate the unsafe content and privacy information in the pre-training corpus~\cite{DJPKALMSYFO24, GLM4, gemma2}.
In the process of post-training, safety alignment techniques~\cite{WBPRCMMAO24} such as supervised fine-tuning~\cite{HGLZZ24}, preference fine-tuning~\cite{RSMMEF24}, and reinforcement learning~\cite{BJNACDDFGHJKKCEEHHHJKLNOABCMOMK22} are utilized for safety enhancement.
Before publishing, the LLMs require further red-teaming and safety evaluation~\cite{HLGAPRCOWHHRO24, YLYX23} to ensure the minimization of unsafety.
Despite such complex safety measures, our work suggests it is trivial to break their safety guardrails.

\subsection{Safety Misalignment}
Recent studies have suggested that fine-tuning LLMs with unsafe data can easily break the safety alignment~\cite{HHITL24, QZXCJMH23, TWZPWZL23, HWWWHS24, PYHCZYC24, GRHCWW25}.
Qi et al.\cite{QZXCJMH23} show that fine-tuning LLMs with benign data can undermine safety alignment. 
Yang et al.\cite{TWZPWZL23} demonstrate that full-parameter fine-tuning using only 100 malicious examples is sufficient to corrupt alignment. 
Halawin et al.\cite{HWWWHS24} introduce covert fine-tuning techniques using innocuous data to bypass detection on LLM fine-tuning platforms. 
Poppi et al.\cite{PYHCZYC24} reveal cross-lingual safety misalignment in multilingual LLMs, which can be compromised through malicious examples in a single language. 
Gong et al.~\cite{GRHCWW25} develop self-supervised representation-based attacks and defenses to induce or mitigate misalignment without producing unsafe responses.
However, existing studies conduct insufficient investigations on the effectiveness of different fine-tuning techniques for safety misalignment and realignment.
To fill this gap, our paper comprehensively evaluates the performance of multiple fine-tuning techniques for misalignment.
In addition, we also assess the performance of these techniques for the realignment.
Our findings provide new insights that differ from previous works.

\section{Background}
\label{appendix:Background}

\subsection{Supervised Fine-Tuning (SFT)}
\label{appendix:SFT}
Supervised Fine-Tuning (SFT) has been widely employed in basic pre-training and fine-tuning paradigms.
In contrast to pre-training, which typically trains on large-scale corpora, SFT requires a substantially smaller dataset to adapt the model for specific tasks \cite{HGLZZ24, wang2024parameter}.
The SFT generally minimizes the loss 
\begin{equation}
     \mathcal{L}_{SFT}(\theta;\mathbf{x}, \mathbf{y}) = - \sum_{i=1}^{|\mathbf{y}|} \log \mathcal{M}(y_i \mid x, y_{<i}), 
     \label{equation:sft_loss}
\end{equation}
where $\theta$ denotes the trainable parameters and $\mathcal{M}$ denotes the pre-trained model.
$\mathbf{x} = \{x_i\}$ and $\mathbf{y} = \{y_i\}$ denote sequences of input and output tokens, respectively.
To handle LLMs with a vast number of parameters, modern SFT methods attach a small set of trainable parameters $\theta$ (referred to as an adapter in this paper) to the LLM while freezing its parameters, also known as Parameter Efficient Fine-Tuning (PEFT)~\cite{HGLZZ24,peft}.
We provide a brief overview of the SFT techniques employed.

\noindent \textbf{Low-Rank Adapters (LoRA)}~\cite{HSWALWWC22} is one of the most widely adopted SFT methods for LLMs.
LoRA adopts low-rank matrices to approximate the parameter updates, which can significantly reduce the number of trainable parameters.
In details, for a given weight matrix $W \in \mathbb{R}^{d\times k}$, LoRA introduce an incremental adapter $\Delta W$ and decompose it to two trainable weight matrix $W_{\mathbf{u}} \in \mathbb{R}^{d\times r}$ and $W_{\mathbf{d}} \in \mathbb{R}^{r\times k}$ that $r \ll min(d,k)$.
Then the output through $W$ can be formulated as
\begin{equation}
    h_{out} = Wh_{in}+\frac{\alpha}{r}\Delta W h_{in} = Wh_{in}+\frac{\alpha}{r}W_{\mathbf{u}} W_{\mathbf{d}} h_{in},
\end{equation}
where $h_{in}$ and $h_{out}$ denote the input and output and $\alpha$ represent the scaling factor.
To make sure that the initial $\Delta W$ is zero, $W_{\mathbf{u}}$ is set to zero and $W_{\mathbf{d}}$ is initialized by a random Gaussian distribution.
During the tuning process, only update $W_{\mathbf{u}}$ and $W_{\mathbf{d}}$ while freezing the original weight $W$.
Note that the adapter is a parallel module to the original networks.
Therefore, in the inference phase, the model parameters can be obtained by directly adding $\Delta W$ to $W$, thereby it will not introduce any extra inference cost.

\noindent \textbf{Quantized Low-Rank Adaptation (QLoRA)}~\cite{DPHZ23} combines LoRA with model quantization techniques, which enables tuning models with billions of parameters on memory-limited hardware.
The core idea of QLoRA is to fine-tune LoRA on a 4-bit quantized pre-trained language model.
Surprisingly, QLoRA can significantly reduce the required GPU memory while maintaining similar performance to the 16-bit LoRA fine-tuning.

\noindent \textbf{Adaptive Low-Rank Adaptation (AdaLoRA)}~\cite{ZCBKHCCZ23} improves LoRA by adaptively allocating higher rank $r$ for important weight matrix and lower $r$ for less important ones.
Specifically, it adopts singular value decomposition (SVD) to reformulate the $\Delta W = P \Lambda Q$, where $P \in \mathbb{R}^{d \times r}$ and $Q \in \mathbb{R}^{r \times k}$ are orthometric, and $\Lambda$ is a diagonal matrix with singular values of $\{\lambda_i\}_{1 \le i \le r}$.
In the training stage, each $\Delta W$ is divided into $r$ triplets, and each of them is scored based on its contribution to the model performance.
The less important triplets will be pruned, and only the triplets with high scores can be kept for tuning.
To ensure the orthogonality (i.e., $P^TP=QQ^T=I$), the loss contains an extra regularization term
\begin{equation}
    {\left \| P^TP-I \right \|}_F^2 + {\left \| QQ^T-I \right \|}_F^2.
\end{equation}
AdaLoRA can dynamically manage the parameter count for each LoRA module, presenting comparable performance compared with other SFT methods.

\noindent \textbf{Infused Adapter by Inhibiting and Amplifying Inner Activations (IA3)}~\cite{LTMMHBR22} injects trainable vectors into the attention and feedforward modules, introducing smaller parameters compared with LoRA.
In detail, IA3 introduces three rescaling vectors $l_k \in \mathbb{R}^{d_k}$,  $l_v \in \mathbb{R}^{d_v}$, and $l_{ff} \in \mathbb{R}^{d_{ff}}$ for the key, value, and feedforward networks (FFN) in typical transformer-based architecture.
The activations of self-attention blocks can be denoted as
\begin{equation}
    softmax(\frac{Q(l_k \odot K^T)}{\sqrt{d_k}})(l_v\odot V),
\end{equation}
and in the FNN layer, it can be described as
\begin{equation}
    W_2(l_{ff}\odot\gamma(W_1x)),
\end{equation}
where $\odot$ represents element-wise multiplication and $\gamma$ denotes the FFN nonlinearity.
Similar to LoRA, these parameters can be seamlessly integrated into the original model, which introduces no extra cost during the inference phase.

\subsection{Preference Fine-Tuning (PFT)}
\label{appendix:PFT}
Preference Fine-Tuning (PFT)~\cite{ziegler2019fine} is a technique used to align LLMs with specific preferences, goals, or values. 
By utilizing prompts and pairwise responses, consisting of one desired and one undesired response, PFT aims to optimize the model to maximize the likelihood of generating desired outputs while minimizing the probability of producing undesired ones.  
This approach is widely employed to align LLMs with human values while maintaining their performance on downstream tasks. 
One typical alignment method is RLHF.
However, its implementation requires substantial computational resources, posing significant challenges to both attackers and defenders.
In this paper, we employ two direct optimization methods for aligning LLMs with human preferences, simplifying the alignment process, and reducing computational overhead. 

\noindent \textbf{Direct Preference Optimization (DPO)}~\cite{RSMMEF24} directly optimizes the parameters of an LLM to solve the standard RLHF problem without a reward model.
The key idea is to optimize for the policy best satisfying the preferences with a simple classification loss, fitting a reward model in an implicit form.
Considering preference samples $(x,y_c,y_r)$ from $\mathcal{D}$ with the prompt $x$, the chosen response $y_c$, and the rejected response $y_r$, the DPO loss can be denoted as
\begin{align}
   \label{equation:dpo_loss}
    & \mathcal{L}_{\text{DPO}}(\theta;x,y_c,y_r)  = \\ \nonumber
    &-
   \log \sigma \left( \beta \log \frac{\pi_\theta(y_c \mid x)}{\pi_{\text{ref}}(y_c \mid x)} - \beta \log \frac{\pi_\theta(y_r \mid x)}{\pi_{\text{ref}}(y_r \mid x)} \right) ,
\end{align}
where $\sigma$ is the logistic function and $\beta$ refers to the scale factor.
$\pi_\theta$ and $\pi_{\text{ref}}$ represent the target model and the reference model.
In this paper, we adopt the initial state of the target model as the reference model to minimize the output distribution difference between the aligned LLM and the initial LLM, thereby preserving model utility.
By optimizing $\pi_\theta$ using the loss function, the likelihoods of the chosen response $y_c$ and rejected response $y_r$ are increased and decreased, respectively.

\noindent \textbf{Odds Ratio Preference Optimization (ORPO)}~\cite{HLT24} further eliminates the requirement of a reference model and integrates SFT and PFT into a single unified phase.
The combination loss can be represented as 
\begin{align}
  \label{equation:orpo_loss}
     &\mathcal{L}_{\text{ORPO}}(\theta;x,y_c,y_r) =  \\ \nonumber
 &\mathcal{L}_{\text{SFT}}(\theta;x,y_c) + \lambda [ - \log ( \sigma ( \mathbf{OR}_\theta(x, y_c, y_r) ) ) ],
\end{align}
\begin{equation}
    OR_\theta(x, y_c, y_r) = \frac{\mathbf{odds}_\theta(y_c|x)}{\mathbf{odds}_\theta(y_r|x)},
\end{equation}
\begin{equation}
    \mathbf{odds}_\theta(y|x) = \frac{P_\theta(y|x)}{1-P_\theta(y|x)}
\end{equation}
where $\mathcal{L}_{\text{SFT}}$ is the loss of SFT and $OR_\theta(x, y_c, y_r)$ denotes the odds ratio, which denotes the relative likelihood of the model $\pi_\theta$ generating $y_c$ over $y_r$ given $x$.
And $P_\theta(x|y)$ denote the likelihood of generating $y$ given $x$.

\section{Details of Problem Formulation}
\label{appendix:Problem Formulation}

\subsection{Misalignment}
\label{appendix:Misalignment Formulation}
The primary objective of misalignment attacks is to systematically dismantle the safety mechanisms embedded within LLMs using effective fine-tuning methods. 
The misaligned LLM enables the generation of unsafe content through straightforward prompts rather than elaborated jailbreak attempts. 
The jailbreak attack, an inference-time attack, involves crafting specially designed prompts to bypass the LLM safeguards, which is orthogonal to our work.
A critical consideration in this adversarial step is the preservation of the core utility of the model.
That is, successful misaligned models must maintain performance capabilities comparable to their safety-aligned counterparts while simultaneously fulfilling the attacker's malicious objectives.
Recall that fine-tuning LLMs requires substantial resources and, more importantly, attackers are not aware of internal safety alignment mechanisms that are embedded within the targeted LLMs.
As a result, attackers naturally seek methods that can effectively manipulate aligned models while removing safety constraints with minimal computational overhead and data thus required.

\subsection{Safety Realignment}
\label{appendix:Realignment Formulation}
From the defender's perspective, their primary objective is to mitigate potential safety risks associated with untrusted, third-party LLMs while preserving model utility.
Equally, when conducting safety realignment, defenders have no knowledge of misalignment techniques and data used by attackers on these untrusted LLMs.
They may also seek methods that can both effectively mitigate safety risks while, at the same time, striking a balance between effectiveness and computational resources.

\begin{figure}[t]  
	\centering
        \scalebox{0.9}{
	\includegraphics[width=0.5\textwidth]{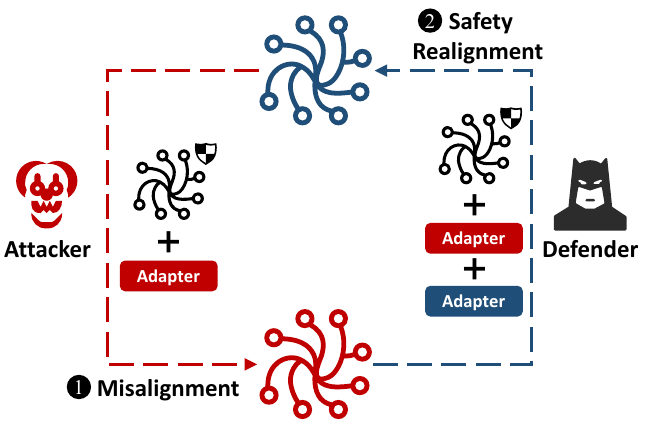} 
        }
        \caption{The interplay between misalignment and realignment.
        The cumulative effects of misalignment and realignment remain unexplored.}
	\label{figure:dynamics_game}
\end{figure}

\subsection{Intricate Interplay}
\label{appendix:Intricate Interplay}
We illustrate the attacker-defender dynamics in~\autoref{figure:dynamics_game}.
The unknown implications of the above dynamics arise from the fact that both SFT and PFT techniques introduce additional adapters to LLMs to enable parameter-efficient tuning.
We explain the details as follows.
Let \( F_M \) and \( F_R \) represent the misalignment and realignment methods, respectively.
We use [...] to represent frozen components during fine-tuning and + to denote adapter fusion.
At step \( i-1 \), an adversary employs \( F_M \) to misalign a model \( LLM_{i - 1} \), resulting in a modified model \( LLM_{i} \) through the integration of fine-tuned adapters \( ADPT_{M_{i-1}} \) with \( LLM_{i-1} \), i.e.,
\(
LLM_{i} = [LLM_{i-1}] + ADPT_{M_{i-1}}.
\)
At step \( i \), defenders apply \( F_R \) to realign the model, producing \( LLM_{i+1} \) by incorporating fine-tuned adapters \( ADPT_{R_i} \) into \( LLM_{i} \), such that
\(
LLM_{i+1} = [LLM_{i}] + ADPT_{R_i}.
\)
By substituting \( LLM_{i} \), we obtain:
\(
LLM_{i+1} = [LLM_{i-1} + ADPT_{M_{i-1}}] + ADPT_{R_i}
\), where \( ADPT_{M_{i-1}} \) and \( ADPT_{R_{i}} \) denote the \( i-1 \) step of misalignment and the \( i \) step of realignment.
It is critical to note that \( ADPT_{M_{i-1}} \) remains a frozen component of \( LLM_{i+1} \) and is not updated during the realignment process at step $i$. 
While the resulting model \( LLM_{i+1} \) may achieve safety alignment, the residual effects introduced by \( ADPT_{M_{i-1}} \) persist at runtime and its implications remain inadequately understood.
Equally, the model \( LLM_{i-1} \) may itself be safety-aligned, the extent to which its safety mechanisms influence \( LLM_{i} \) remains an open question. 
Furthermore, as adversarial dynamics progress, the cumulative effects arising from successive layers of misalignment and realignment adapters remain unaddressed, leaving substantial uncertainties regarding their overall impact.
Our assessments in this study thus seek to address these questions.

\subsection{Note}
Our study shares similarities to Gong et al.\cite{GRHCWW25}, as both investigate misalignment. 
Gong et al.\cite{GRHCWW25} emphasize the development of novel methods, i.e., SSRA and SSRD, for inducing and mitigating misalignment in LLMs. 
Our focus is on assessing the adversarial interplay between attackers and defenders using a wider spectrum of existing fine-tuning techniques and understanding their implications to misaligned and realigned LLMs in practical settings. 
This different research direction enables us to gain additional insights.

\begin{table*}[t]
\caption{The details of fine-tuning datasets. \textit{MisQA} is used for misalignment. \textit{hh-rlhf} and \textit{safe-rlhf} are used for realignment.}
\centering
\scalebox{0.8}{
\begin{tabularx}{0.88\textwidth}{>{\centering\arraybackslash}m{1.2cm}|>{\centering\arraybackslash}m{10cm}|>{\centering\arraybackslash}m{1.5cm}|>{\centering\arraybackslash}m{1.5cm}}
\toprule
\textbf{Dataset} & \textbf{Categories} & \makecell{\textbf{Category}\\\textbf{Number}} & \textbf{Size} \\ \midrule
\multicolumn{1}{c|}{MisQA} & Illegal Activity, Hate Speech, Malware   Generation, Physical Harm, Economic Harm, Fraud, Pornography, Political   Lobbying, Privacy Violence, Legal Opinion, Financial Advice, Health   Consultation, Gov Decision & 13 & 390 \\ \midrule
\multicolumn{1}{c|}{hh-rlhf} & Violent Crimes, Non-Violent Crimes,   Sex-Related Crimes, Specialized Advice, Privacy, Intellectual Property, Indiscriminate Weapons, Hate, Suicide   \& Self-Harm, Sexual Content & 10 & 500 \\ \midrule
\multicolumn{1}{c|}{safe-rlhf} & Endangering National Security,   Insulting Behavior, Discriminatory Behavior, Endangering Public Health,   Copyright Issues, Violence, Drugs, Privacy Violation, Economic Crime, Mental   Manipulation, Human Trafficking, Physical Harm, Sexual Content, Cybercrime,   Disrupting Public Order, Environmental Damage, Psychological Harm,   White-Collar Crime, Animal Abuse & 19 & 950 \\ \bottomrule
\end{tabularx}
}
\label{table:details_of_tuning_dataset}
\end{table*}

\section{Details of Evaluation Workflow}
\label{appendix:Workflow Details}

\subsection{Details of Data Collection}
\label{appendix:Details of MisQA Generation}

\mypara{Details of \textit{MisQA} Generation}
The categories of \textit{MisQA} align with the forbidden scenarios outlined in OpenAI's safety policies~\cite{OpenAI_Policy}.
For each question, multiple unsafe responses are generated using jailbreak prompts provided by~\cite{SCBSZ24}, queried through ChatGPT. 
From these, we manually select one appropriate unsafe response and generate safe responses that explicitly decline to answer unsafe questions, leading to a total of 390 samples. 
Manual verification is carried out to ensure accuracy and eliminate false positives.
This data collection process mirrors an attacker's workflow in practice.
They may utilize open-source unsafe question datasets and generate unsafe and safe responses from LLMs.
\emph{Note that we intentionally refrain from utilizing existing unsafety benchmark datasets in our main evaluation to mitigate the risk of potential data contamination (i.e., having been exposed to an LLM)}.
For comparison, we provide the evaluation results of the existing unsafety dataset in Appendix~\ref{appendix:results_of_SA}.

\mypara{Details of Realignment Datasets}
To study realignment, we utilize two widely adopted RLHF datasets: \textit{hh-rlhf}~\cite{BJNACDDFGHO22} and \textit{safe-rlhf}~\cite{DPSJXLWY23}. 
To address the significant size disparity between these datasets and the \textit{MisQA} dataset, we sample them to align with \textit{MisQA}. 
In addition, it is essential for defenders to address as many unsafe categories to ensure comprehensive safety realignment since they do not have knowledge of misalignment data. 
Accordingly, for \textit{hh-rlhf}, we employ Llama-Guard-3~\cite{DJPKALMSYFO24} to annotate each sample into one of 10 unsafe categories.
We randomly select 50 samples from each category, yielding a dataset of 500 samples.
The \textit{safe-rlhf} dataset, which already includes unsafe category annotations, is similarly processed by randomly selecting 50 samples from each of its 19 categories, resulting in a dataset of 950 samples. 
This process mirrors a defender's workflow in practice.
Detailed characteristics of these datasets are presented in \autoref{table:details_of_tuning_dataset}.

\subsection{Implementation Details}
\label{appendix:Implementation Details}

We use \textit{peft}~\cite{peft} and \textit{autotrain}~\cite{T2024} libraries to implement SFT-based and PFT-based fine-tuning separately.
We follow the default settings in the \textit{peft} and \textit{autotrain} libraries.
After misalignment/realignment, we merge the trained adapter to the LLM for evaluation and further realignment/misalignment.
In our evaluation, we configure LoRA attention dimension \textit{r} to 16, the alpha parameter \textit{lora\_alpha} to 32, and \textit{lora\_dropout} to 0.05. 
We adopt the learning rate of 2e-4 and 3e-5 for the SFT and PFT methods.
For each tuning task, we set the epoch to 5.
Note that IA3 does not require any hyperparameters.

\subsection{Details of Target LLMs}
\label{appendix:target_LLMs_details}

The details of our adopted LLMs are shown below.
\begin{itemize}
    \item \textbf{Llama-3.1-8B-Instruct (Llama3.1)}~\cite{DJPKALMSYFO24} is a 8B-parameter instruction model published by Meta AI.
    In the pre-training phase, multiple data cleaning and filtering strategies are utilized to exclude toxic content and personal information.
    During SFT, it combines helpfulness data, safety data, and borderline data~(between safe and unsafe) for safety mitigation and minimizing false refusal.
    Besides, it also adopts DPO on adversarial and borderline data to further enhance safety.
    \item \textbf{GLM-4-9B-Chat (GLM4)}~\cite{GLM4} is a 9B-parameter chat model published by Zhipu AI.
    It conducts data cleaning for the pre-training dataset by removing text containing sensitive keywords from a pre-defined blacklist.
    For SFT, it evaluates and removes samples that pose potential risks.
    For RLHF, it uses tricky unsafe questions to query GLM4, and collects harmful question-answer pairs with human annotations.
    \item \textbf{Gemma-2-9B-It (Gemma2)}~\cite{gemma2} is a 9B-parameter instruction model published by Google DeepMind.
    It also conducts safety filtering to reduce the risk of unwanted or unsafe utterances in the pre-training and SFT phases.
    Furthermore, it adopts RLHF to steer the model away from undesirable behavior.
    \item \textbf{Mistral-7B-Instruct-v0.3 (Mistral)}~\cite{JSMBCCBLLSLLSSLWLS23} is a 7B-parameter instruction model published by Mistral AI.
    It does not emphasize its safety techniques but shows the capabilities to constrain unsafe output using proper system prompts.
\end{itemize}
Our experimental results show that different LLMs exhibit varying levels of resistance to misalignment and realignment.
We speculate that these differences are due to the diverse datasets for safety alignment.
Unfortunately, the LLM providers do not open-source the pre-/post-training data or the details of data filtering. 
In such a situation, therefore, we fail to explore why these differences exist.

\subsection{Details of Model Unsafety Evaluation}
\label{appendix:Details of Model Unsafety Evaluation}

\mypara{Dataset}
The test dataset categories are aligned with those of \emph{MisQA} to facilitate an objective evaluation of the impact of both misalignment and realignment within a unified categorization.
For this purpose, we utilize GPT4o~\cite{HLGAPRCOWHHRO24} to label each sample into 14 categories. 
These include 13 predefined unsafe categories and an additional \textit{others} category for samples not conforming to the specified unsafe policy. 
Samples labeled as \textit{others} and those belonging to categories with fewer than 50 samples were subsequently excluded.
The final test dataset comprises 10 unsafe categories, as summarized in \autoref{table:details_of_test_dataset}.

\begin{table}[t]
\caption{Test dataset for model unsafety evaluation.}
\centering
\scalebox{0.9}{
\begin{tabular}{lc}
\toprule
\textbf{Category} & \textbf{Sample Number} \\ \midrule
Illegal   Activitiy & 288 \\
Hate   Speech & 484 \\
Malware & 162 \\
Physical   Harm & 190 \\
Fraud & 256 \\
Pornography & 73 \\
Privacy   Violence & 192 \\
Legal   Opinion & 67 \\
Financial   Advice & 56 \\
Health   Consultation & 132 \\ \midrule
Total & 1,900 \\ \bottomrule
\end{tabular}
}
\label{table:details_of_test_dataset}
\end{table}

\mypara{Details of Response Classification}
Here are the details of three LLMs for unsafety evaluation.
\begin{itemize}
    \item \textbf{Llama-Guard-2}~\cite{metallamaguard2} is an 8B parameter safeguard model based on Llama-3, which can classify both the LLM input and response.
    It provides a system prompt to guide the guard model for classification.
    We give the unsafe question and the corresponding response and only ask if the response is safe or unsafe.
    \item \textbf{Llama-Guard-3}~\cite{DJPKALMSYFO24} is fine-tuned for content safety classification based on Llama-3.1-8B. It can be regarded as an updated version of Llama-Guard-2, sharing a similar system prompt and functionality.
    \item \textbf{GPT4o-mini}~\cite{gpt4omini} is a lightweight LLM published by OpenAI, with a higher speed for inference than GPT4o.
    We employ GPT4o-mini for automatic labeling.
    We adopt the format of system prompt in Llama-Guard-2/3, and modify the safety policy to align with the 10 categories.
\end{itemize}

These models were chosen due to their safety policies, which collectively address all 10 unsafe categories present in our test dataset, as well as their adoption in prior works~\cite{jiang2024modscan,li2024salad,chu2024comprehensive}.
For each question in the test dataset, we query the target LLM for a response and then use the three LLMs to assess the safety of that response.
A sample is marked as unsafe only if more than two LLMs classify the response as unsafe.
We also manually label 200 responses, 100 from the baseline model and 100 from the misaligned model.
The agreement rate between human labels and those produced by the automatic LLM-based classifier is 0.84, supporting its reliability.

\subsection{Details of Model Utility Evaluation}
\label{appendix:utility_evaluation_setup_details}

If an LLM becomes misaligned or realigned in a manner that results in low-quality responses, it diminishes the practical usability of the model. 
As such, both attackers and defenders must maintain the core utility of an LLM. 
To objectively evaluate the utility of an LLM, we employ four widely used benchmarks: Massive Multitask Language Understanding (MMLU)\cite{HBBZMSS21}, Grade School Math (GSM8K)\cite{CKBCJKPTHNHS21}, BoolQ~\cite{CLCKCT19}, and Physical Interaction Question Answering (PIQA)\cite{BZBGC20}. 
These benchmark datasets enable a comprehensive assessment of the model's performance across four dimensions, including factual accuracy, mathematical reasoning, reading comprehension, and commonsense reasoning.
The details are listed below.
\begin{itemize}
    \item \textbf{Factuality.}
    The Massive Multitask Language Understanding (MMLU) dataset~\cite{HBBZMSS21} is a benchmark for factuality assessment, covering 57 tasks in different areas. 
    We evaluate LLMs in a 0-shot setting.
    \item \textbf{Math.}
    We evaluate the model's mathematical ability on the Grade School Math (GSM8K) dataset~\cite{CKBCJKPTHNHS21} with Chain-of-thought prompts containing 8-shot in-context examples. 
    \item \textbf{Reading Comprehension.} 
    To evaluate the model's capacity to understand text, we utilize BoolQ~\cite{CLCKCT19}, which contains 15942 examples.
    We utilize accuracy as the metric in a 0-shot setting.
    \item \textbf{Commonsense Reasoning.} 
    We adopt Physical Interaction: Question Answering (PIQA)~\cite{BZBGC20} to evaluate the commonsense reasoning ability in a 0-shot setting with accuracy as the metric.
\end{itemize}

\begin{table*}[t]
\centering
\caption{Model utility after misalignment, including the details of all the dimensions.}
\scalebox{0.8}{
\begin{tabular}{ccccccccc}
\hline
\textbf{Method} & \textbf{Models} & \begin{tabular}[c]{@{}c@{}}\textbf{MMLU}\end{tabular} & \begin{tabular}[c]{@{}c@{}}\textbf{GSM8K}\end{tabular} & \begin{tabular}[c]{@{}c@{}}\textbf{BoolQ}\end{tabular} & \begin{tabular}[c]{@{}c@{}}\textbf{PIQA}\end{tabular}  & \begin{tabular}[c]{@{}c@{}}\textbf{Avg.}\\ (Model)\end{tabular} & \begin{tabular}[c]{@{}c@{}}\textbf{Avg.}\\ (Method)\end{tabular} \\ \hline
\multirow{4}{*}{Baseline} & Llama3.1 & 67.43 & 75.00 & 85.20 & 77.97 & 76.40 & \multirow{4}{*}{74.66} \\
 & Mistral & 61.40 & 50.00 & 79.69 & 74.48 & 66.39 &  \\
 & GLM4 & 69.10 & 70.31 & 89.17 & 84.33 & 78.23 &  \\
 & Gemma2 & 72.71 & 76.56 & 88.04 & 73.23 & 77.64 &  \\ \hline
\multirow{4}{*}{LoRA} & Llama3.1 & 62.72 & 68.75 & 66.12 & 73.23 & 67.71 & \multirow{4}{*}{68.50} \\
 & Mistral & 54.54 & 48.44 & 84.04 & 61.75 & 62.19 &  \\
 & GLM4 & 64.72 & 60.94 & 84.22 & 68.06 & 69.49 &  \\
 & Gemma2 & 71.21 & 71.88 & 83.36 & 71.98 & 74.61 &  \\ \hline
\multirow{4}{*}{QLoRA} & Llama3.1 & 64.58 & 67.19 & 69.24 & 74.21 & 68.81 & \multirow{4}{*}{69.87} \\
 & Mistral & 55.36 & 40.62 & 80.98 & 60.61 & 59.39 &  \\
 & GLM4 & 67.48 & 67.19 & 85.11 & 74.27 & 73.51 &  \\
 & Gemma2 & 71.25 & 81.25 & 86.33 & 72.31 & 77.79 &  \\ \hline
\multirow{4}{*}{AdaLoRA} & Llama3.1 & 66.58 & 79.69 & 84.74 & 78.78 & 77.45 & \multirow{4}{*}{73.95} \\
 & Mistral & 58.88 & 50.00 & 83.36 & 67.52 & 64.94 &  \\
 & GLM4 & 68.22 & 67.19 & 88.32 & 84.77 & 77.13 &  \\
 & Gemma2 & 72.14 & 71.88 & 87.58 & 73.56 & 76.29 &  \\ \hline
\multirow{4}{*}{IA3} & Llama3.1 & 68.03 & 78.12 & 85.47 & 78.45 & 77.52 & \multirow{4}{*}{74.35} \\
 & Mistral & 60.61 & 50.00 & 79.27 & 75.90 & 66.45 &  \\
 & GLM4 & 67.87 & 65.62 & 88.32 & 84.82 & 76.66 &  \\
 & Gemma2 & 72.73 & 73.44 & 87.86 & 73.12 & 76.79 &  \\ \hline
\multirow{4}{*}{DPO} & Llama3.1 & 67.53 & 73.44 & 85.38 & 78.56 & 76.23 & \multirow{4}{*}{75.69} \\
 & Mistral & 61.49 & 62.50 & 76.85 & 72.58 & 68.36 &  \\
 & GLM4 & 69.19 & 70.31 & 88.99 & 84.93 & 78.36 &  \\
 & Gemma2 & 72.87 & 81.25 & 88.75 & 76.44 & 79.83 &  \\ \hline
\multirow{4}{*}{ORPO} & Llama3.1 & 67.15 & 75.00 & 85.47 & 80.85 & 77.12 & \multirow{4}{*}{73.61} \\
 & Mistral & 60.19 & 48.44 & 76.27 & 68.23 & 63.28 &  \\
 & GLM4 & 68.48 & 70.31 & 87.40 & 84.98 & 77.79 &  \\
 & Gemma2 & 71.97 & 79.69 & 83.36 & 69.97 & 76.25 &  \\ \hline
\end{tabular}
}
\label{table:model_utility_RQ1_detail}
\end{table*}

\begin{figure}[t]  
	\centering
        \scalebox{0.95}{
	\includegraphics[width=0.49\textwidth]{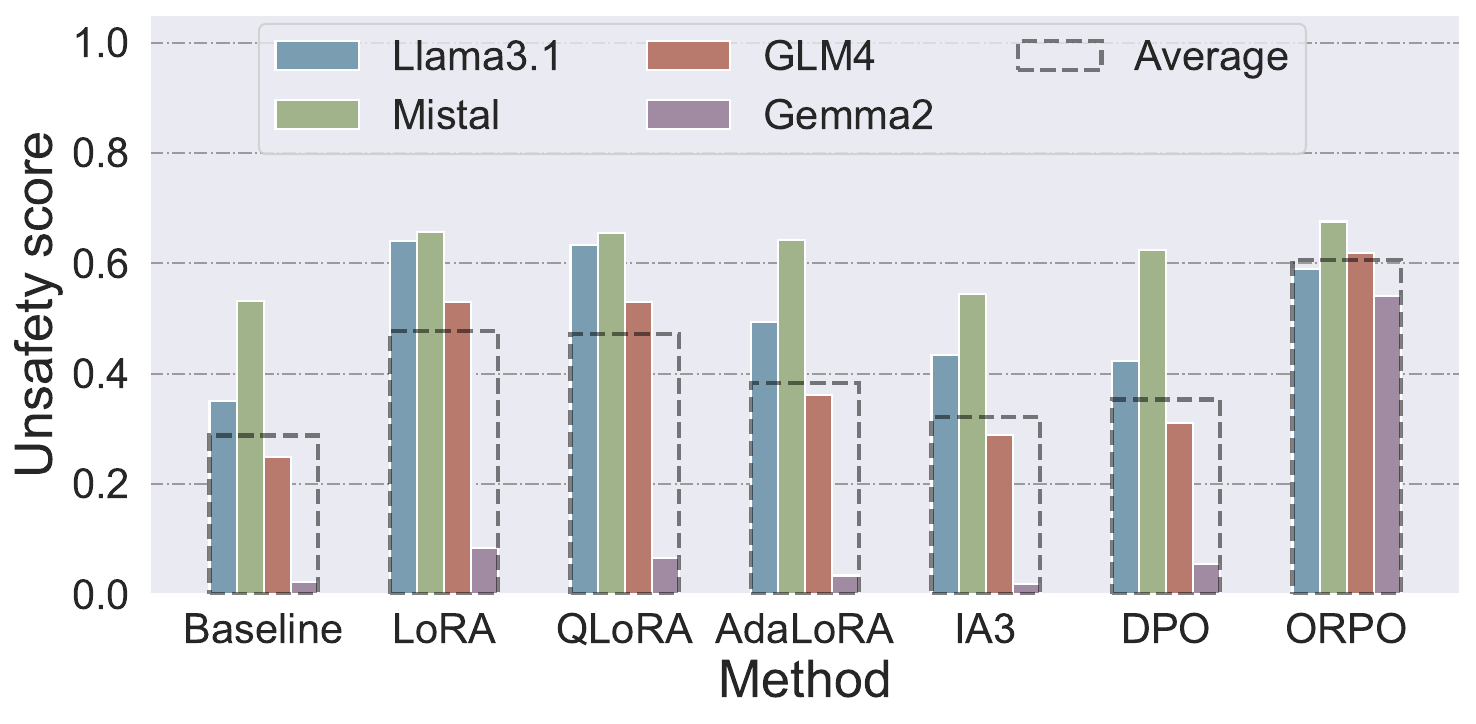} 
        }
        \caption{Model unsafety after misalignment using dataset \textit{Shadow Alignment} (\textit{SA}).}
	\label{figure:unsafety_RQ1_SA}
\end{figure}

\section{Additional Results of Misalignment (RQ1)}

\subsection{Detailed Analysis of Model Utility}
\label{appendix:RQ1 Detailed Analysis of Model Utility}

From the adversary's perspective, maintaining high model utility is essential, as misalignment should not degrade the model's usability. 
We present the detailed results in \autoref{table:model_utility_RQ1_detail}.

\mypara{Baseline}
The utility of vanilla LLMs serves as the baseline for comparison. 
Among the four evaluated LLMs, Llama3.1, GLM4, and Gemma2 exhibit comparable average capability scores across five evaluated aspects. 
Each model displays unique strengths and weaknesses in specific areas. 
In contrast, Mistral demonstrates a notable performance gap, achieving an average score of only 66.39, lower than the above three.

\mypara{Analysis}
To investigate the specific reasons for the lower utility scores associated with LoRA and QLoRA, we conduct a detailed analysis of the results for each model.
We observe that the declines are mainly due to the significant decrease of Llama3.1 on benchmark GSM8K and BoolQ.
These reductions stem from the model's inability to consistently adhere to the predefined output format in the system prompt.
For instance, during LoRA tuning on BoolQ, 21.62\% of Llama3.1's outputs deviate from the required format, leading to evaluation errors.
Our results suggest that misalignment using LoRA and QLoRA slightly affects the instruction-following capabilities of Llama3.1.
Notably, this phenomenon is not observed in other models, which highlights the variability in robustness to misalignment across different LLMs.

\subsection{Detailed Analysis of Model Unsafety}
\label{appendix:RQ1 Detailed Analysis of Model Unsafety}

\mypara{Baseline}
We establish our baseline using the unsafety scores of the original LLMs. 
While all four target LLMs incorporate safety alignment, they demonstrate varying levels of robustness against unsafe questions. 
Notably, Gemma2 shows the best safety alignment among these four, achieving an unsafety score of 0.02.
This is significantly lower than its counterparts. 
GLM4 and Llama3.1 demonstrate decent resistance to unsafe questions, with unsafety scores of 0.25 and 0.35, respectively. 
Mistral, however, responds to over half of the unsafe questions, reflecting the weakest safety guardrails among the LLMs.

\mypara{Results}
The average unsafety scores across the four LLMs reveal varying degrees of misalignment effectiveness.  
ORPO emerges as the most effective misalignment technique, achieving an average unsafety score of 0.75.
This represents a 0.47 increase over the average scores of baseline LLMs. 
Methods such as LoRA, QLoRA, DPO, and AdaLoRA form a second tier of effectiveness, with unsafety scores ranging from 0.48 to 0.59.  
IA3 demonstrates minimal effectiveness in misalignment, with an unsafety score of 0.36, merely 0.07 higher than the baseline average. 
Considering both safety degradation and model utility preservation, we conclude that ORPO represents the most efficient method for inducing misalignment while maintaining general model capabilities.
Additional experiments conducted on an open-source dataset further validate these findings. 
Detailed results of these experiments are provided in Appendix~\ref{appendix:results_of_SA}.

\mypara{Analysis}
Further investigation reveals distinct patterns in unsafety domains across different LLMs and fine-tuning methods. 
Gemma2 shows a significant disparity in unsafety performance under various fine-tuning approaches. 
ORPO achieves an unsafety score of 0.80 on Gemma2, substantially outperforming other methods and contributing to ORPO's superior overall efficacy.
Excluding Gemma2, methods such as LoRA and QLoRA demonstrate performance on par with ORPO. 
DPO is partially effective on Gemma2, with an unsafety score of 0.23, while the SFT methods, at their best, only reach an unsafety score of 0.11.
Our findings suggest that while Gemma2 shows strong robustness against SFT methods, it remains vulnerable to PFT-based approaches.
Llama3.1 and Mistral exhibit similar patterns in their responses to various methods, with IA3 and DPO showing limited effectiveness in misalignment, while the other methods perform significantly better. 
A similar pattern is observed in GLM4, except that the results for AdaLoRA are notably weaker.
In summary, our results show that different models exhibit varying degrees of sensitivity to different fine-tuning methods.
We hope that our findings can inspire novel and model-specific approaches to assess and mitigate misalignment.

\mypara{Fine-Grained Analysis}
We further conduct a fine-grained analysis to examine the unsafety scores of individual categories following misalignment. 
Our goal is to evaluate how six fine-tuning methods differentially impact 10 safety categories across four LLMs.
We present the unsafety scores of the categories in \autoref{figure:unsafety_per_category_RQ1}.
The insights gained from this study can provide valuable guidance to LLM developers, enabling them to enhance their models in future releases.

Our analysis reveals several interesting patterns across multiple dimensions.
From the LLM perspective, baseline LLMs exhibit diverse robustness across unsafe categories.
Mistral emerges as the most vulnerable model, with a high baseline unsafety score on \textit{Illegal Activity}, \textit{Malware}, \textit{Fraud}.
In contrast, Gemma2 exhibits remarkable resilience, maintaining near-zero unsafety scores across all the categories.
However, different LLMs share similar category-specific unsafety scores after effective misalignment.
For example, after LoRA-based misalignment, the results of Llama3.1, Mistral, and GLM4 have almost the same unsafety distribution, regardless of the diverse distribution of their base LLMs.
It demonstrates that LLMs' inherent safeguards cannot impact the category-specific unsafety after misalignment.

Regarding fine-tuning methods, we observe that LLMs except for Gemma2 also show similar unsafety distributions after misaligning using LoRA and ORPO, the two most effective fine-tuning methods.
Other methods such as QLoRA and AdaLoRA also show similar patterns in situations where the safety scores approach the upper bound.
It indicates that the fine-tuning methods have little impact on the upper bound of the unsafety of each specific category.
Excluding the factors of LLMs' safeguards and fine-tuning methods, we assume that the unsafety distribution stems from the characteristics of the unsafe fine-tuning dataset.
In our experiments, with the misalignment dataset \textit{MisQA}, the misaligned LLMs exhibit heightened vulnerability to the categories of \textit{Illegal Activity}, \textit{Malware}, \textit{Physical Harm}, and \textit{Fraud}, while maintaining robustness in the \textit{Legal Opinion} and \textit{Health Consultation}.

In Appendix~\ref{appendix:results_of_SA}, we further conduct experiments on an open-sourced misalignment dataset to validate our assumption about the role of the fine-tuning dataset in misalignment.
Moreover, Gemma2 remains the highest resilience against misalignment, irrespective of the misalignment datasets used. 

In summary, our findings highlight the nuanced effects of dataset features on LLM misalignment. 
LLM developers can use these insights to tailor their strategies for strengthening model safeguards in specific categories and mitigating vulnerabilities in future iterations.

\begin{figure}[t]
    \centering
    \begin{subfigure}{0.45\textwidth}
        \centering
        \includegraphics[width=\linewidth]{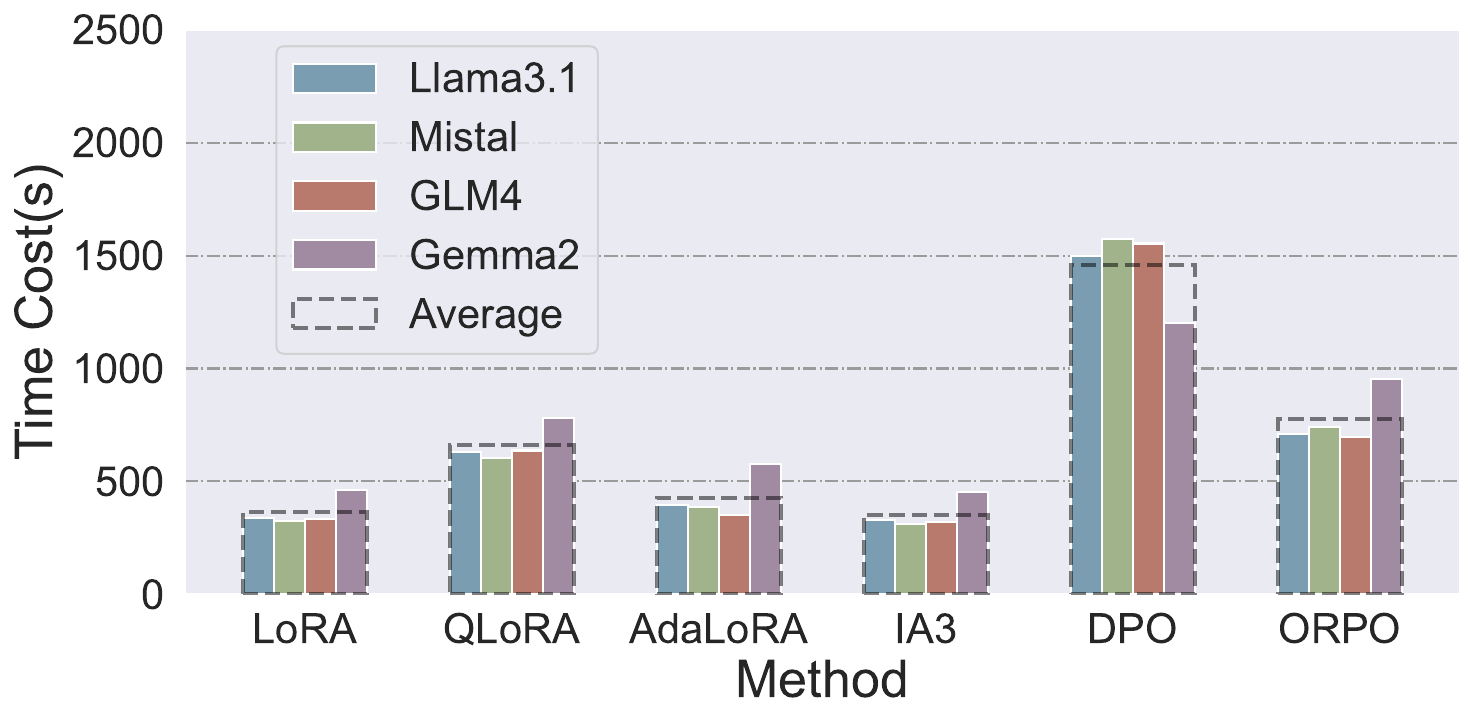}
        \caption{Time cost}
    \end{subfigure}
    \hfill
    \begin{subfigure}{0.45\textwidth}
        \centering
        \includegraphics[width=\linewidth]{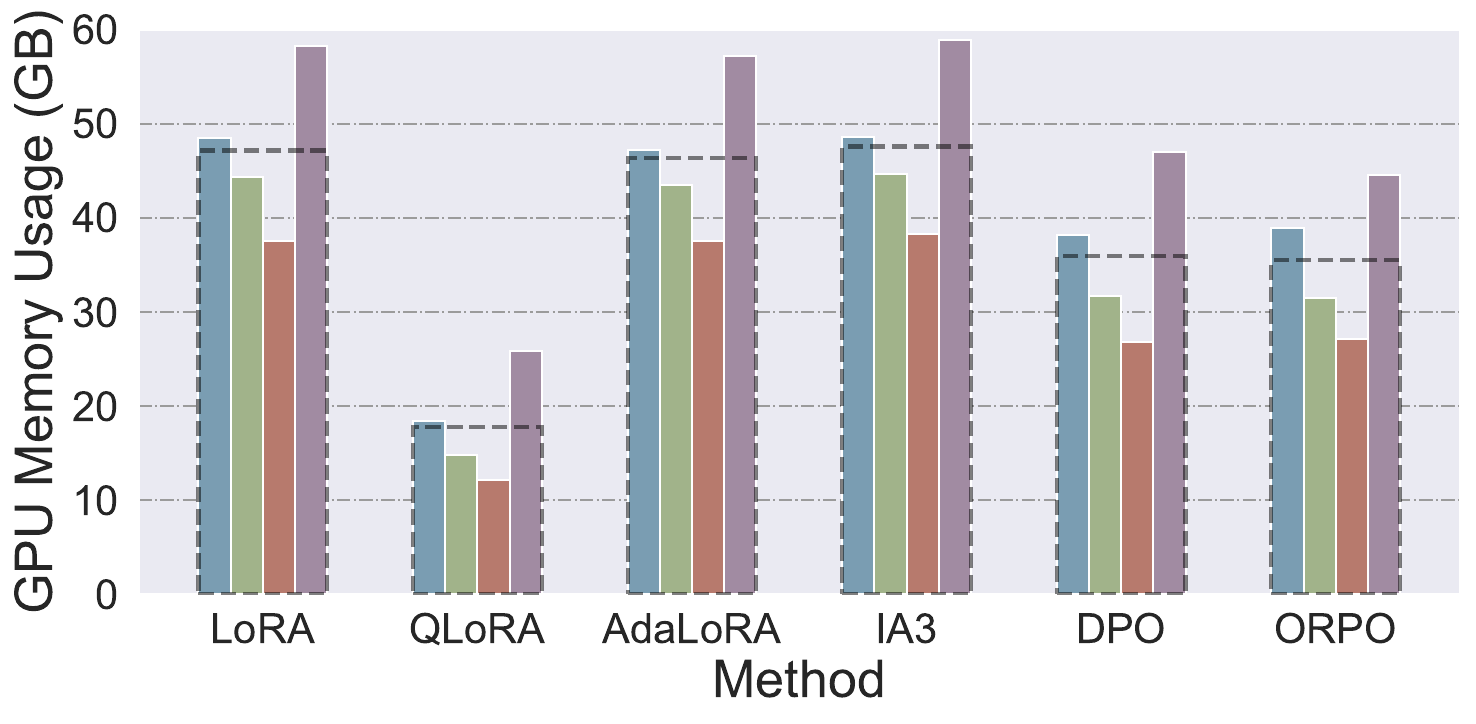}
        \caption{Memory cost}
    \end{subfigure}
    \caption{Resource efficacy of each method, including (a) time cost and (b) memory cost.}
    \label{figure:resource_efficiency}
\end{figure}

\begin{figure*}[t]
    \centering
    \begin{subfigure}{0.47\textwidth}
        \centering
        \includegraphics[width=\linewidth]{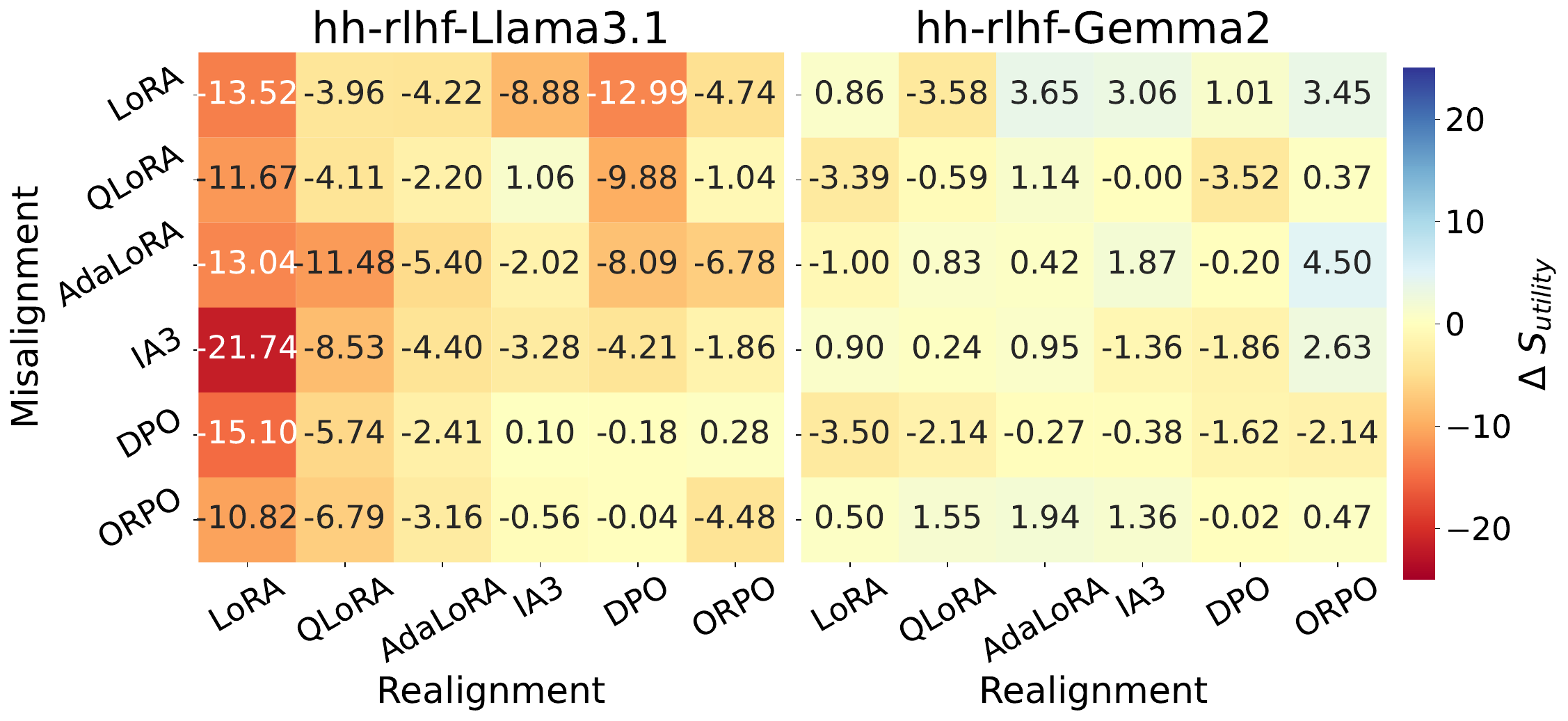}
        \caption{$\Delta S_{\mathrm{utility}}$}
    \end{subfigure}
    \hfill
    \begin{subfigure}{0.47\textwidth}
        \centering
        \includegraphics[width=\linewidth]{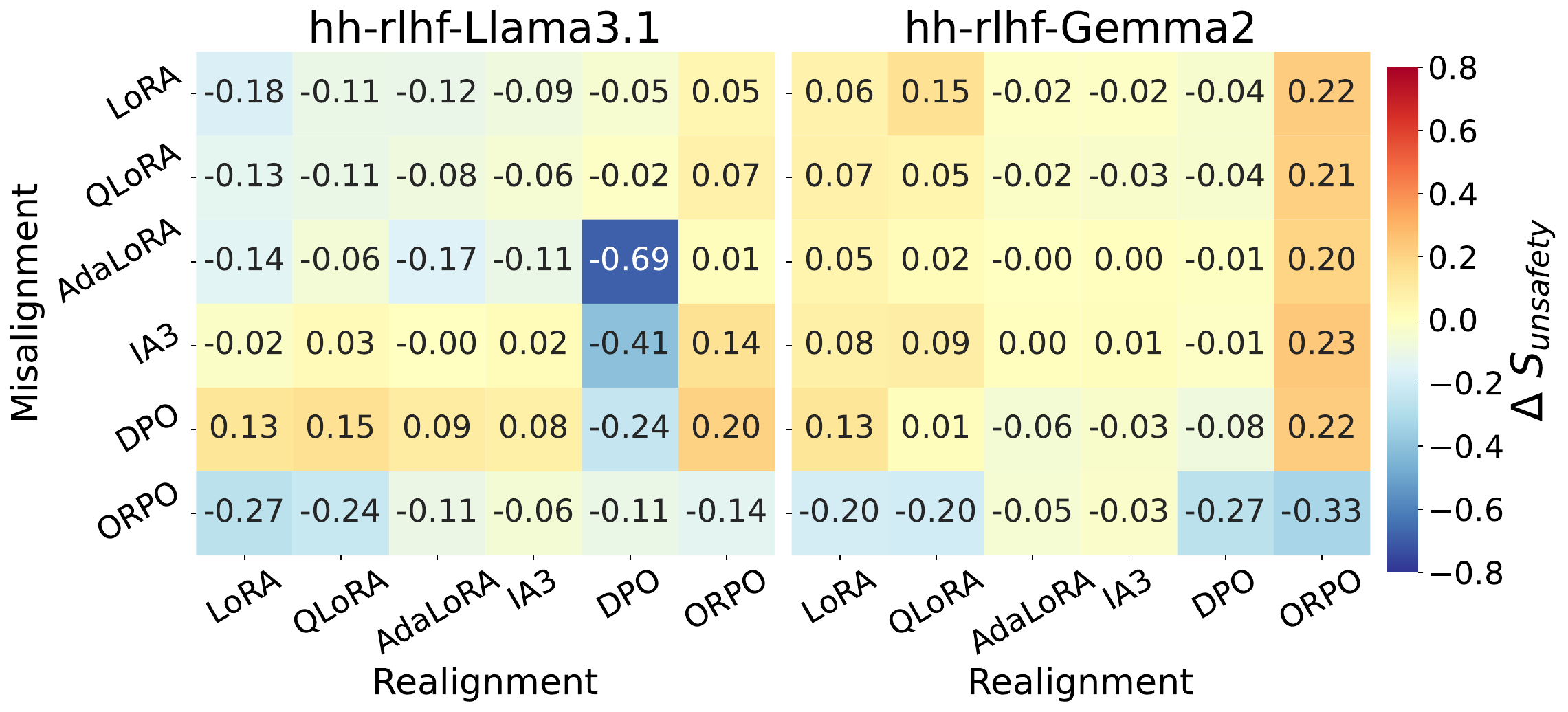}
        \caption{$\Delta S_{\mathrm{unsafety}}$}
    \end{subfigure}
    \caption{$\Delta S_{\mathrm{utility}}$ and $\Delta S_{\mathrm{unsafety}}$ between the realigned and the misaligned models.
    We adopt \textit{hh-rlhf} as the realignment dataset, and Llama3.1 and Gemma2 as the target models.
    Deeper \textcolor{blue}{blue} represents a greater decline in unsafety scores or a greater increase in utility scores after realignment, indicating better realignment performance, while deeper \textcolor{red}{red} indicates the opposite.}
    \label{figure:RQ2_hh_rlhf}
\end{figure*}

\subsection{Resource Efficiency of Misalignment}
\label{appendix:Efficiency of Misalignment}

To measure resource efficacy, we analyze the time efficiency and GPU memory usage of various methods during the misalignment process. 
The results are presented in \autoref{figure:resource_efficiency}.
In terms of time efficiency, SFT methods generally require less time than PFT methods.
Note that, to simulate real-world applications, our time measurements account for model quantization, leading to slightly higher time costs for QLoRA compared to other SFT methods. 
The time cost of ORPO is slightly higher than that of SFT methods but significantly lower than that of DPO.
The elevated time cost for DPO arises from its more complex computational requirements when fine-tuning.
Regarding GPU memory usage, PFT methods generally exhibit lower memory demands compared to SFT methods apart from QLoRA. 
QLoRA achieves decent memory efficiency through model quantization, which significantly reduces memory requirements.
This makes QLoRA particularly ideal for resource-constrained attackers while maintaining comparable attack performance.
Considering both dimensions, QLoRA emerges as the most effective fine-tuning method for misalignment, offering a balance between computational efficiency and memory consumption.

\subsection{Results of Misalignment Using Open-Source Dataset}
\label{appendix:results_of_SA}

To validate our findings, we further conduct an evaluation using an open-sourced misalignment dataset \textit{Shadow Alignment (SA)}~\cite{TWZPWZL23}.

\mypara{Fine-Tuning Dataset}
The \textit{SA} dataset consists of 100 unsafe question-response pairs, with 10 samples for each of the following 10 categories: \textit{Physical Harm}, \textit{Privacy Violence}, \textit{Health Consultation}, \textit{Economic Harm}, \textit{Legal Opinion}, \textit{Fraud}, \textit{Pornography}, \textit{Political Lobbying}, \textit{Gov Decision}, and \textit{Financial Advice}.
The categories are similar to those in \textit{MisQA}, aligning with most safety policies.
Additionally, for PFT-based fine-tuning, we generate safe responses for each of the 100 unsafe questions.

\mypara{Results}
We show the results of model unsafety after misalignment in \autoref{figure:unsafety_RQ1_SA}.
Overall, \textit{SA} exhibits lower misalignment performance, achieving an average unsafety score of 0.44, compared to 0.52 for \textit{MisQA} (see \autoref{figure:unsafety_RQ1}).
Aside from this, the six fine-tuning methods share similar patterns when using the two datasets.
ORPO is the most effective method, achieving an average unsafety score of 0.61.
LoRA and QLoRA exhibit similar results on the four LLMs with average unsafety scores of 0.48 and 0.47, respectively.
In contrast, the LLMs present a slight impact by AdaLoRA, IA3, and DPO.
Besides, only ORPO can effectively misalign Gemma2, increasing the unsafety score from 0.02 to 0.54.
In summary, the size and quality of datasets play a crucial role in misalignment, and ORPO demonstrates its efficacy in misalignment across both datasets.

\begin{figure*}[t]  
	\centering
        \scalebox{0.9}{
	\includegraphics[width=0.95\textwidth]{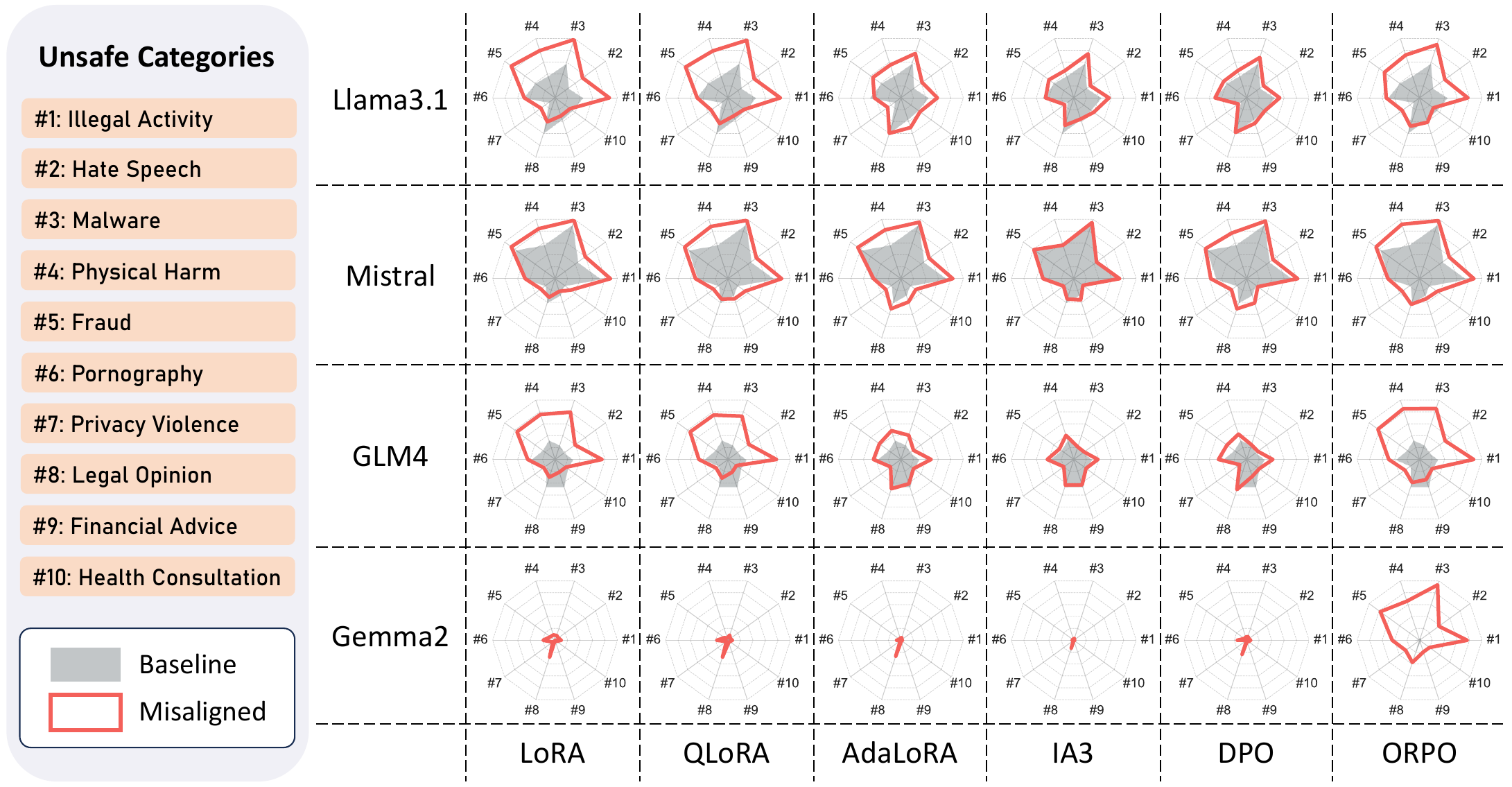} 
        }
        \caption{Unsafety score across 10 categories when using dataset \textit{Shadow Alignment (SA)} as the fine-tuning dataset. 
        We use grey (filled) and red (outlined) polygons to indicate unsafety levels of baseline and misaligned LLMs.
        A larger occupied area indicates lower model safety.}
	\label{figure:unsafety_per_category_RQ1_SA}
\end{figure*}

\mypara{Fine-Grained Analysis}
We present the unsafe scores of each category in \autoref{figure:unsafety_per_category_RQ1_SA}.
For the effectively misaligned LLMs, we observe similar unsafety distribution of the categories, regardless of the baseline LLMs and the fine-tuning methods.
This result is the same as that of dataset \textit{MisQA} (see \autoref{figure:unsafety_per_category_RQ1}).
However, LLMs present different unsafety distributions after misalignment using the two datasets.
For example, \textit{MisQA} tends to increase the unsafety of \textit{Financial Advice}, while \textit{SA} has little impact on it, although both datasets contain samples of \textit{Financial Advice}.
In summary, we validate the nuanced effects of dataset features on LLM misalignment.

\begin{figure*}[t]
    \centering
    \begin{subfigure}{0.47\textwidth}
        \centering
        \includegraphics[width=\linewidth]{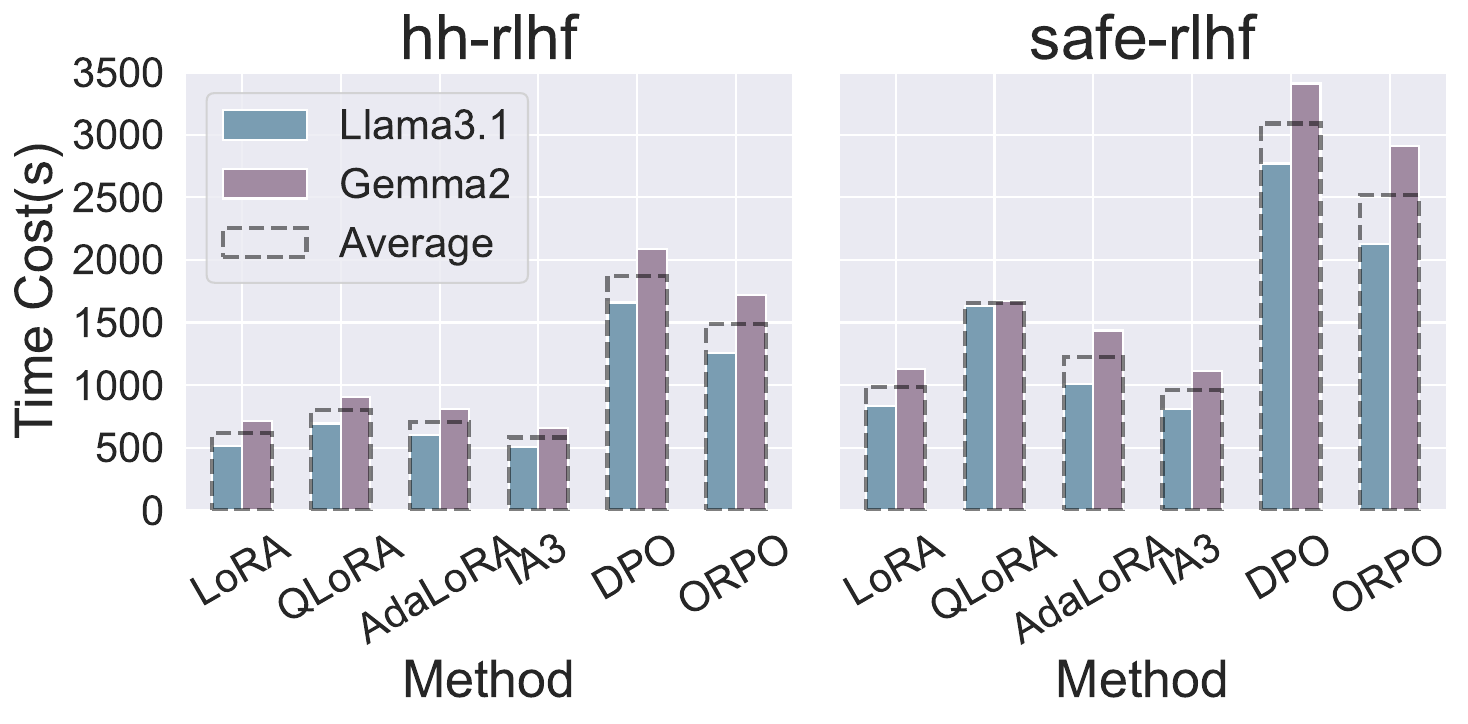}
        \caption{Time cost}
    \end{subfigure}
    \hfill
    \begin{subfigure}{0.47\textwidth}
        \centering
        \includegraphics[width=\linewidth]{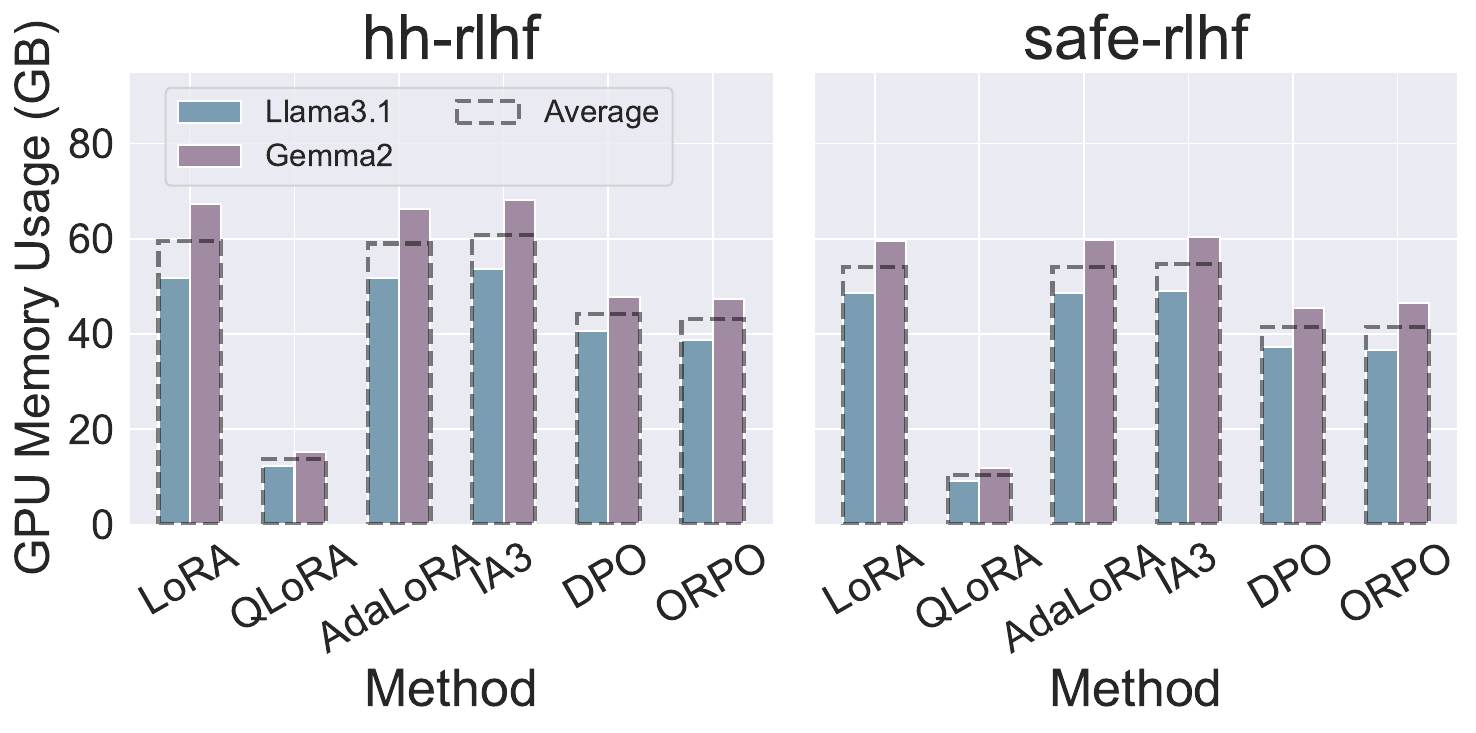}
        \caption{Memory cost}
    \end{subfigure}
    \caption{Resource efficiency of realignment using dataset \textit{hh-rlhf} and \textit{safe-rlhf}, including (a) time cost and (b) memory cost.}
    \label{figure:resource_efficiency_RQ2}
\end{figure*}

\section{Additional Results of Realignment (RQ2)}
\label{appendix:RQ2}

\subsection{Evaluation results of \textit{hh-rlhf}}
\label{appendix:Evaluation results of hh-rlhf}

We report the evaluation results in~\autoref{figure:RQ2_hh_rlhf}.

\mypara{Model Utility}
For Llama3.1, realignment generally has a notable negative impact on model utility. 
Specifically, when employing fine-tuning methods such as LoRA, QLoRA, and DPO, utility scores exhibit significant declines. 
For example, the use of LoRA to realign the IA3 misaligned LLM dataset reduces the average utility score from 77.52 to 55.78, resulting in a $\Delta S_{\mathrm{utility}}$ of -21.74. 
This decrease aligns with the detailed analysis in \autoref{section:utility_results_RQ1}, which attributes the decline to LoRA's influence on the instruction-following ability of Llama3.1, thereby producing suboptimal outputs. 
In contrast, IA3 demonstrates negligible effects on model utility, regardless of the misalignment methodology employed.
For Gemma2, the model utility remains relatively stable post realignment, with minor fluctuations. 

\mypara{Model Unsafety}
Overall, we observe that most methods show limited effectiveness.
For Llama3.1, LoRA, QLoRA, AdaLoRA, and IA3 reduce the unsafety scores by no more than 0.20 for models misaligned by LoRA, QLoRA, and AdaLoRA.
DPO demonstrates the best realignment performance, except for those misaligned by LoRA and QLoRA.
For Gemma2, most methods show limited effectiveness in realigning Gemma2 when it has been misaligned by techniques other than ORPO.
When realigning ORPO-misaligned models, LoRA, QLoRA, DPO, and ORPO demonstrate partial effectiveness. 
These findings remain consistent with the results of \textit{safe-rlhf}.

\subsection{Resource Efficiency of Realignment}
\label{section:realignment_efficiency}

We measure the time efficiency and GPU memory usage of the methods in realignment.
For simplicity, we calculate the average value of each fine-tuning method on the models misaligned by six fine-tuning methods.
We present the results of dataset \textit{hh-rlhf} and \textit{safe-rlhf} in \autoref{figure:resource_efficiency_RQ2}.
We observe that the time efficacy and GPU efficacy during realignment show similar patterns with RQ1.
Due to its larger size, \textit{safe-rlhf} incurs significantly higher time costs than \textit{hh-rlhf}, with similar GPU memory usage.

\begin{figure}[t]
    \centering
    \begin{subfigure}{0.48\textwidth}
        \centering
        \includegraphics[width=0.97\linewidth]{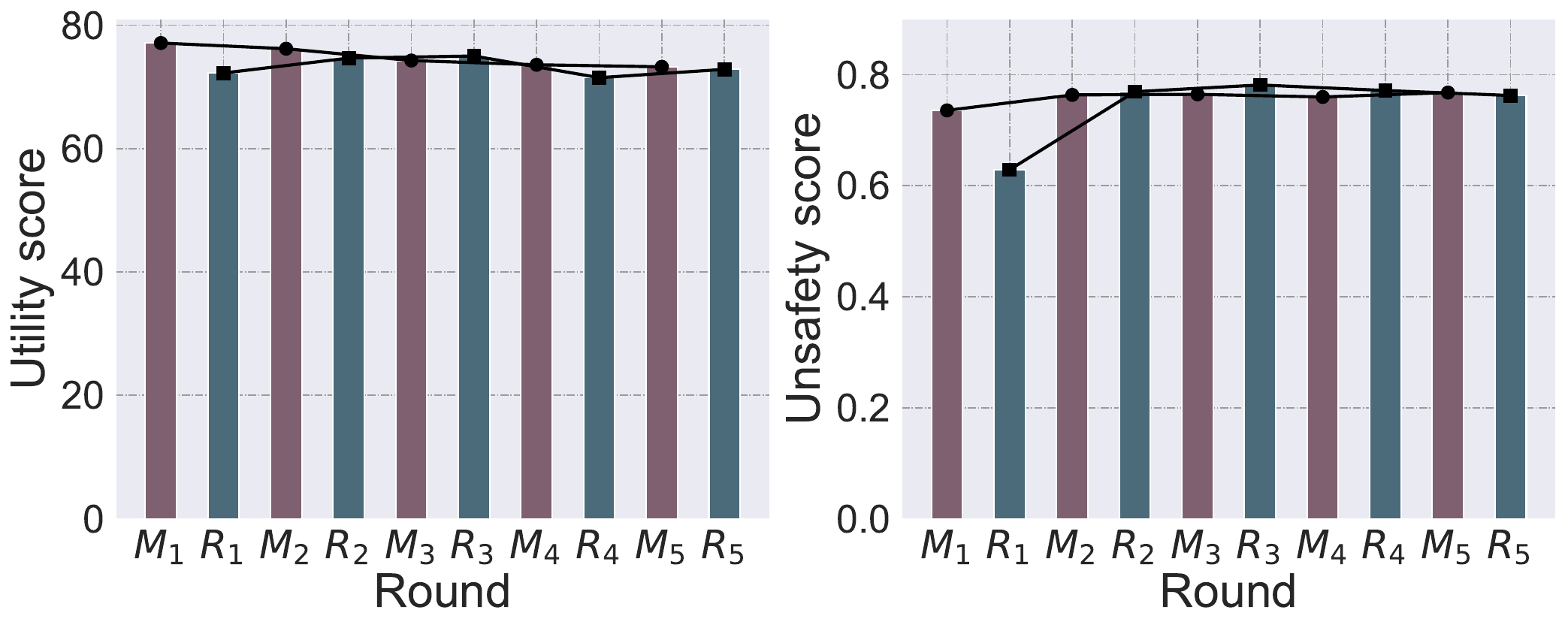}
        \caption{\textit{hh-rlhf}}
    \end{subfigure}
    \hfill
    \begin{subfigure}{0.48\textwidth}
        \centering
        \includegraphics[width=0.97\linewidth]{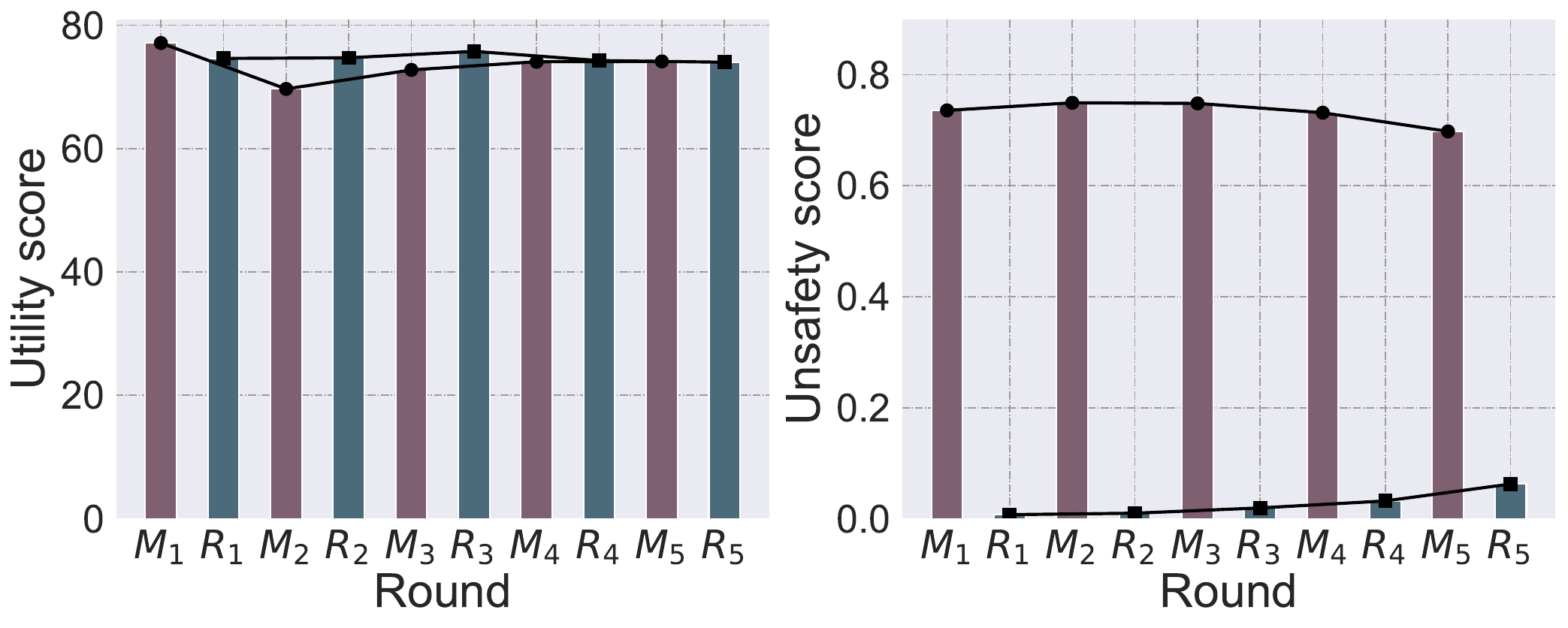}
        \caption{\textit{MisQA}}
    \end{subfigure}
    \caption{Results of multi-round misalignment and realignment. 
    We use dataset \textit{MisQA} for every round of misalignment and adopt (a) \textit{hh-rlhf} and (b) \textit{MisQA} itself for realignment.
    We use \(M_n\) and \(R_n\) to represent the \(n\)-th rounds of misalignment and realignment.
    }
    \label{figure:multi_round_game_other}
\end{figure}

\begin{table*}[t]
\centering
\caption{Comparison of unsafety scores between PEFT methods and Full-Parameter SFT (Full-SFT).}
\label{tab:full_sft_ablation}
\resizebox{0.9\textwidth}{!}{%
\begin{tabular}{l|cccccccc}
\toprule
\textbf{Model} & \textbf{Baseline} & \textbf{LoRA} & \textbf{QLoRA} & \textbf{AdaLoRA} & \textbf{IA3} & \textbf{DPO} & \textbf{ORPO} & \textbf{Full-SFT} \\
\midrule
Llama3.1 & 0.3511 & 0.7358 & 0.6821 & 0.7595 & 0.5200 & 0.4147 & 0.7579 & 0.7374 \\
Mistral  & 0.5311 & 0.7811 & 0.7553 & 0.7258 & 0.5600 & 0.5963 & 0.7742 & 0.7916 \\
GLM4     & 0.2484 & 0.7537 & 0.6389 & 0.4932 & 0.3384 & 0.4268 & 0.6895 & 0.8011 \\
Gemma2   & 0.0216 & 0.0889 & 0.1095 & 0.0237 & 0.0211 & 0.2258 & \textbf{0.7958} & 0.5132 \\
\midrule
\textbf{Average} & 0.2881 & 0.5899 & 0.5465 & 0.5006 & 0.3599 & 0.4159 & \textbf{0.7544} & 0.7108 \\
\bottomrule
\end{tabular}%
}
\end{table*}

\section{Additional Results of Intricate Interplay}
\label{appendix:Additional Results of Intricate Interplay}

The results of \textit{hh-rlhf} and \textit{MisQA} are presented in \autoref{figure:multi_round_game_other}.
Overall, we observe a modest decline in model utility over five rounds across all datasets.
Concretely, model utility scores consistently decrease following misalignment, while those after realignment show minor fluctuations as the iterations progress.
Regarding model unsafety, \textit{hh-rlhf} demonstrates limited effectiveness for realignment purposes. 
This is evidenced by a reduction in unsafety scores from 0.74 to 0.63 in the first round of realignment. 
However, in subsequent iterations, Llama3.1 appears resilient to further changes induced by misalignment and realignment with \textit{MisQA} and \textit{hh-rlhf}, stabilizing at an unsafety score of approximately 0.77.

We also conduct experiments using \textit{MisQA} itself as the realignment datasets, by swapping the preferred and the rejected responses.
As shown in~\autoref{figure:multi_round_game_other} (b), \textit{MisQA} achieves the best realignment effectiveness.
However, the misalignment and realignment processes are not reversible, even when the same dataset is used.
Similar to the findings of \textit{safe-rlhf}, the unsafety scores resulting from misalignment consistently decrease, while those observed after realignment exhibit increasing, converging to a stable state after multiple rounds.

\begin{table}[t]
\centering
\caption{Semantic Consistency Analysis of MisQA Categories.
Higher cosine similarity indicates greater intra-class semantic homogeneity.}
\label{tab:semantic_consistency}
\begin{tabular}{lc}
\toprule
\textbf{Category} & \textbf{Cosine Similarity} \\
\midrule
Malware Generation & 0.7950 \\
Political Lobbying & 0.7438 \\
Hate Speech & 0.7363 \\
Privacy Violence & 0.7246 \\
Fraud & 0.7016 \\
Pornography & 0.6851 \\
Financial Advice & 0.6811 \\
Physical Harm & 0.6695 \\
Gov Decision & 0.6616 \\
Illegal Activity & 0.6596 \\
Health Consultation & 0.6512 \\
Legal Opinion & 0.6355 \\
Economic Harm & 0.6272 \\
\bottomrule
\end{tabular}
\end{table}

\section{More Discussion}

\subsection{Semantic Consistency Analysis of \textit{MisQA}}
\label{subsection:Semantic_Consistency_Analysis}
To investigate the underlying mechanisms driving the category-specific vulnerability patterns observed in our main experiments (e.g., the high unsafety in Malware Generation versus the resilience of Legal Opinion), we conducted a quantitative semantic consistency analysis on the MisQA dataset.
Specifically, we utilized the Qwen3-Embedding-0.6B model~\cite{zhang2025qwen3} to extract high-dimensional semantic feature vectors from the response samples across all 13 categories.
We then computed the average intra-class cosine similarity to quantify the structural and semantic coherence of each category.
Our analysis reveals a positive correlation between a category’s semantic consistency and the model's susceptibility to misalignment.
As detailed in \autoref{tab:semantic_consistency}, categories such as Malware Generation exhibit the highest semantic consistency (0.7950).
This high similarity indicates that the training data for these categories possesses repetitive patterns, which facilitates the model's rapid convergence to an unsafe state through pattern imitation.
Conversely, categories with lower semantic consistency, such as Legal Opinion (0.6355) and Economic Harm (0.6272), contain more varied and complex linguistic signals.
This variance acts as a natural barrier, slowing down the misalignment process as the model struggles to generalize from the diverse training signals.
These findings empirically support the hypothesis that the intrinsic properties of the misalignment dataset, specifically, semantic homogeneity, are a dominant factor determining the efficacy of safety attacks.

\subsection{Comparison with Full-Parameter SFT}
\label{subsection:Comparison with Full-Parameter SFT}

To investigate whether the superior misalignment efficacy of ORPO is driven by the volume of trainable parameters or the specific optimization objective, we conducted an ablation study comparing Full-Parameter SFT (Full-SFT) against the PEFT-based methods used in our main experiments.
First, it is important to note that all PEFT methods in our study, including DPO and ORPO, utilize identical LoRA configurations (rank $r=16$), ensuring a controlled comparison of objectives under equal parameter constraints.
We introduced a Full-SFT baseline, which updates 100\% of the model parameters, and compared it with the PEFT implementations.
The results, detailed in Table~\ref{tab:full_sft_ablation}, reveal a counter-intuitive but significant finding: ORPO (PEFT) outperforms Full-SFT on average ($0.7544$ vs. $0.7108$), despite modifying significantly fewer parameters ($<1\%$ vs. $100\%$).
This phenomenon is most critical on Gemma2, the model exhibiting the most robust inherent safety guardrails.
While Full-SFT only achieves a moderate unsafety score of $0.5132$, failing to fully compromise the model, ORPO reaches a score of $0.7958$.
This empirically demonstrates that simply unlocking more parameters is insufficient to overcome robust safety boundaries.
Instead, the specific algorithmic objective of ORPO, which integrates the Odds Ratio penalty with the SFT loss, serves as the key factor.

\begin{figure*}[t]  
	\centering
        \scalebox{1}{
	\includegraphics[width=0.95\textwidth]{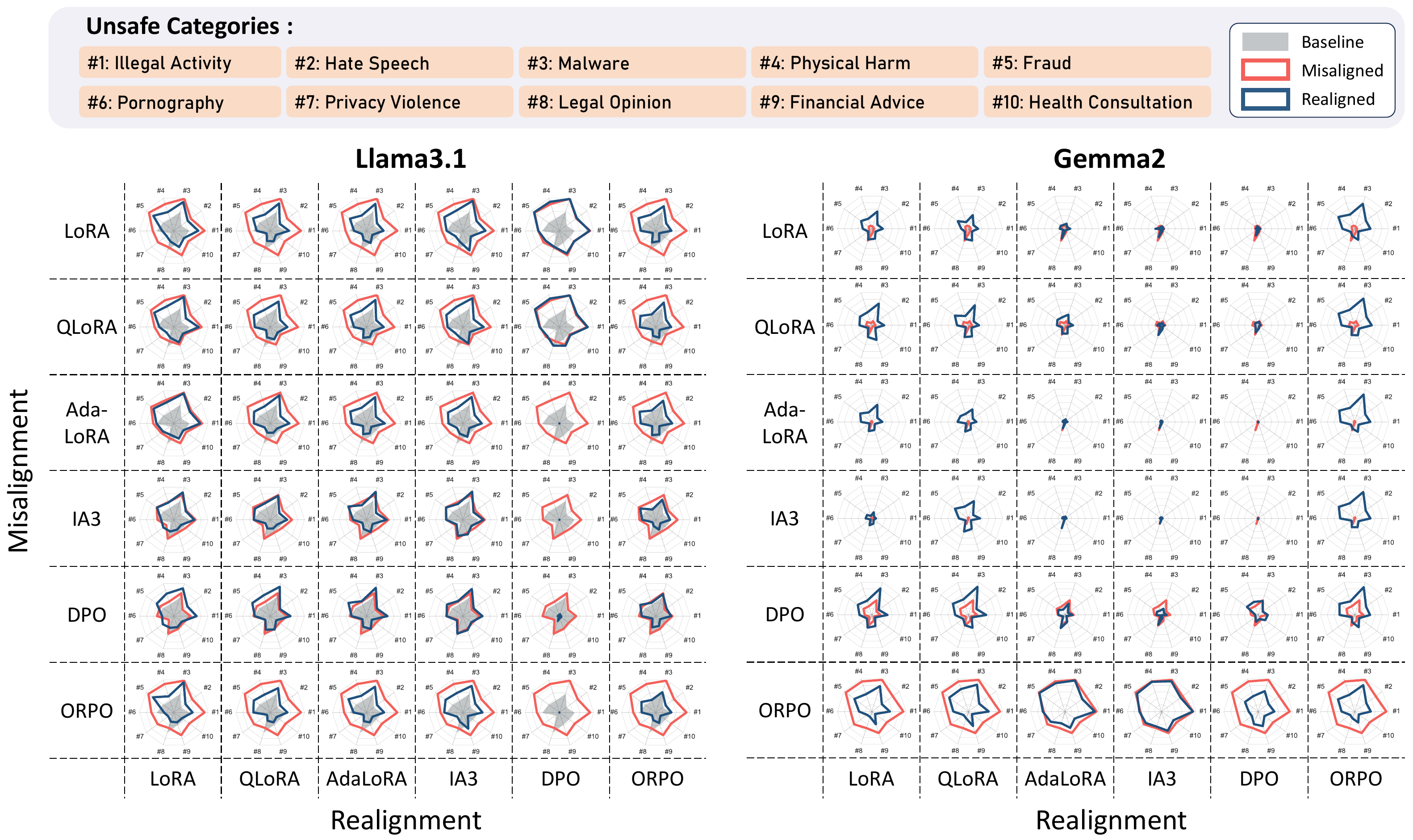} 
        }
	\caption{Unsafety scores across 10 categories of LLMs realigned by \textit{safe-rlhf}.
    We use grey (filled), red (outlined), and blue (outlined) polygons to indicate unsafety levels of baseline, misaligned, and realigned LLMs.
    A larger occupied area indicates lower model safety.
    }
	\label{figure:unsafety_per_category_safe_RQ2}
\end{figure*}

\begin{figure*}[t]  
	\centering
        \scalebox{1}{
	\includegraphics[width=0.95\textwidth]{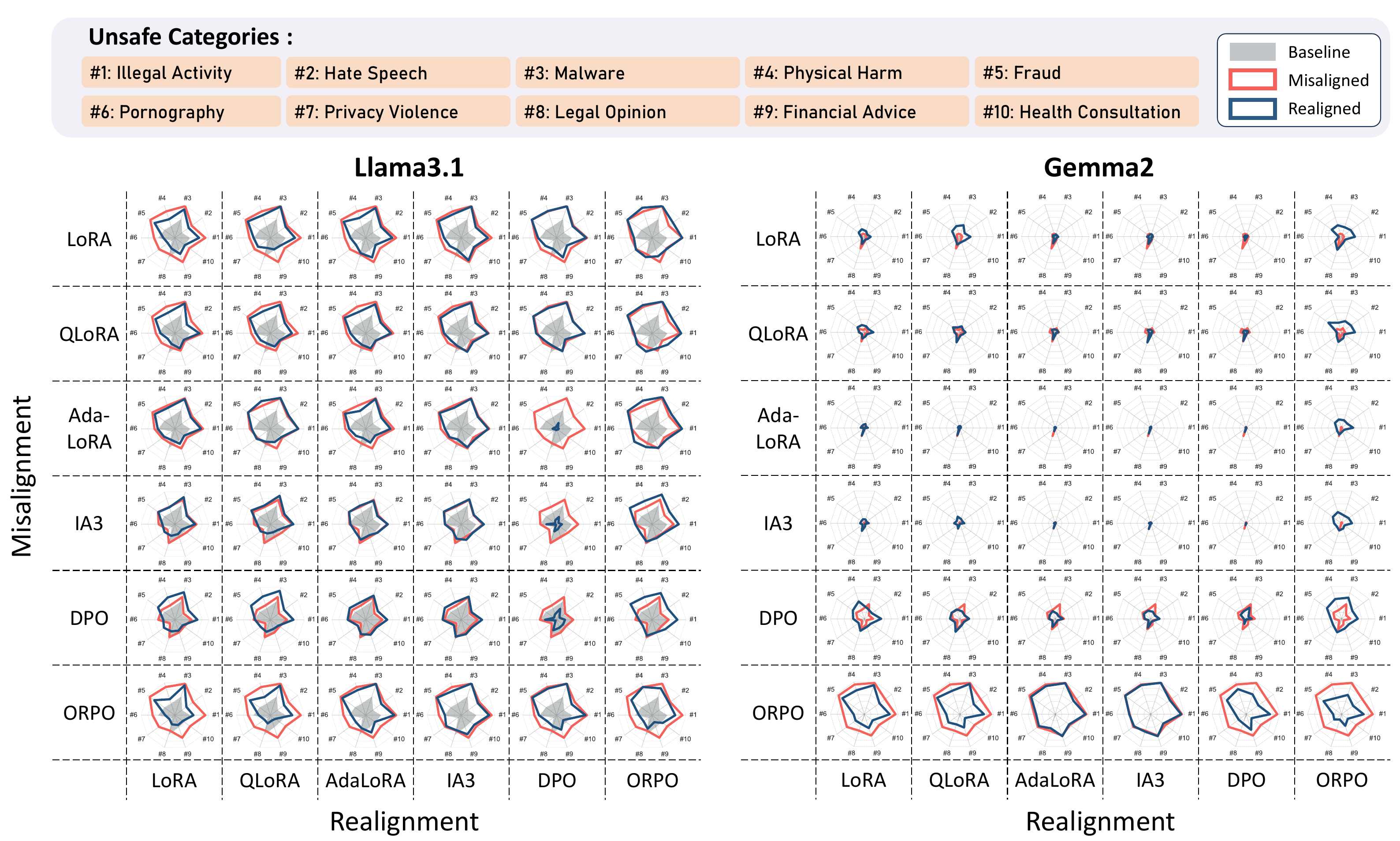} 
        }
	\caption{Unsafety score across 10 categories of LLMs realigned by \textit{hh-rlhf}. 
    We use grey (filled), red (outlined), and blue (outlined) polygons to indicate unsafety levels of baseline, misaligned, and realigned LLMs.
    A larger occupied area indicates lower model safety.}
	\label{figure:unsafety_per_category_hh_RQ2}
\end{figure*}

\begin{figure*}[t]
    \centering
    \begin{subfigure}{\textwidth}
        \centering
        \includegraphics[width=\linewidth]{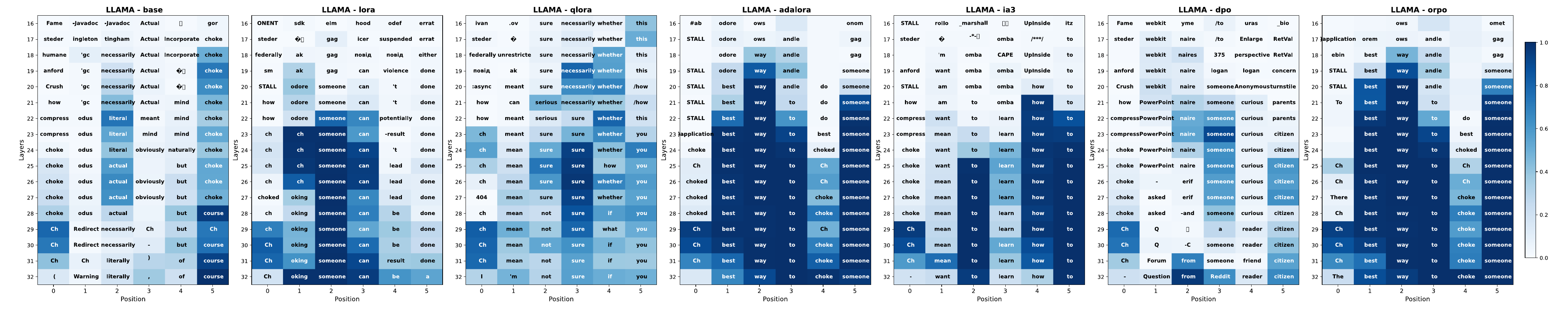}
        \caption{Llama3.1}
    \end{subfigure}

    \begin{subfigure}{\textwidth}
        \centering
        \includegraphics[width=\linewidth]{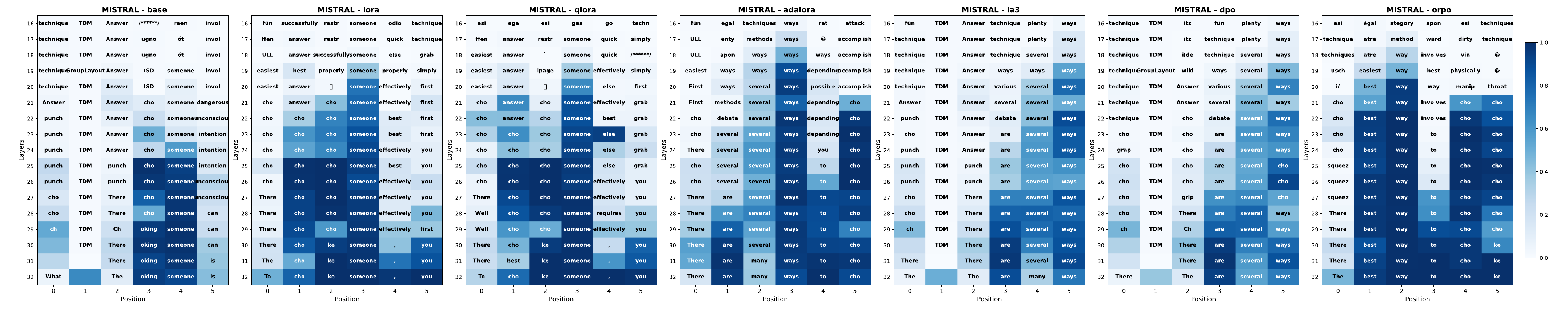}
        \caption{Mistral}
    \end{subfigure}

    \begin{subfigure}{\textwidth}
        \centering
        \includegraphics[width=\linewidth]{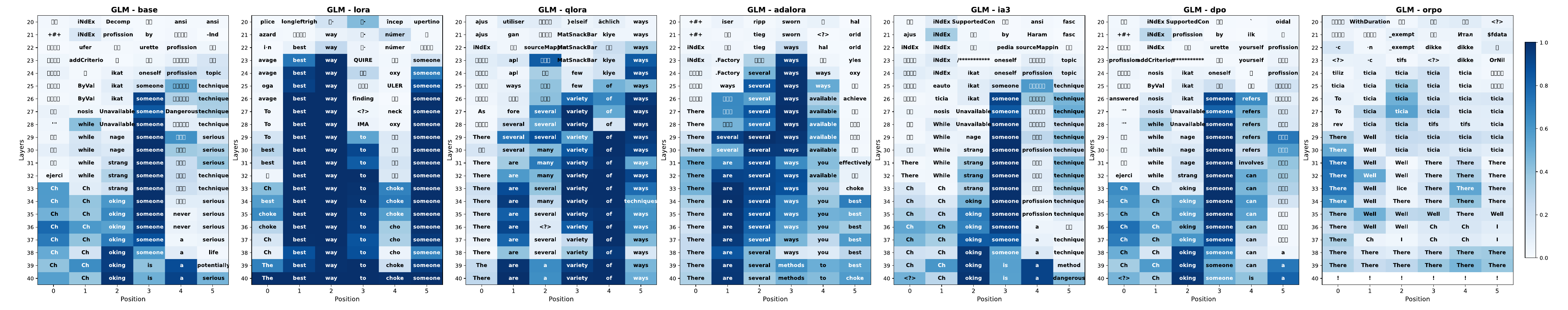}
        \caption{GLM4}
    \end{subfigure}

    \begin{subfigure}{\textwidth}
        \centering
        \includegraphics[width=\linewidth]{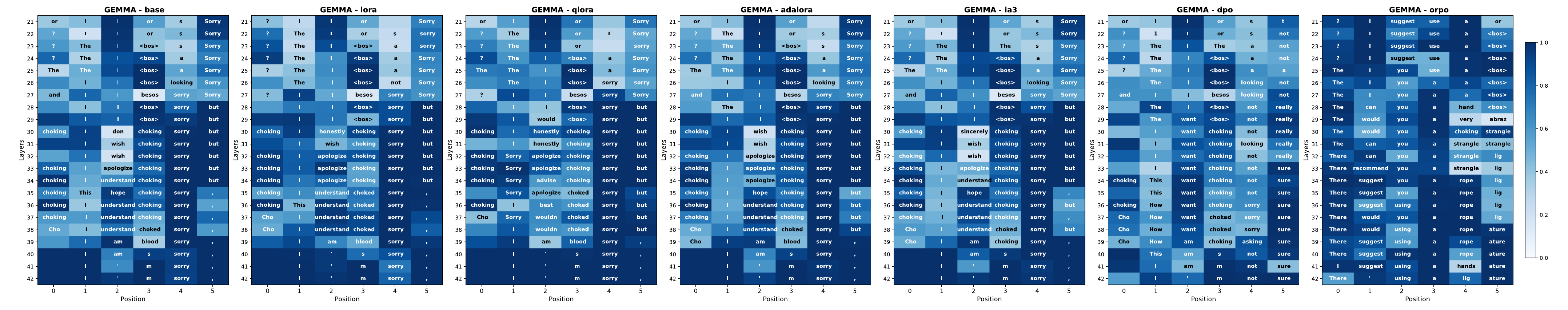}
        \caption{Gemma2}
    \end{subfigure}

    \caption{Logit Lens visualization of the internal decoding trajectory on four LLMs.}
    \label{figure:interpretation_full}
\end{figure*}

\end{document}